\documentclass[pteplogo, dvips]{ptephy_v1}
\usepackage{amsmath,amssymb}
\usepackage{mathrsfs}
\usepackage{bm}

\preprintnumber{CHIBA-EP-218}

\begin{document}
\title{Gluon bound state and asymptotic freedom derived from the Bethe--Salpeter equation}

\author{Hitoshi Fukamachi}
\affil{Kaisei Senior and Junior High Schools, Tokyo 116-0013, Japan \email{\rm{fukamachi-ht@kaiseigakuen.jp}}}
\author{Kei-Ichi Kondo}
\affil{Department of Physics, 
Chiba University, Chiba 263-8522, Japan 
\email{\rm{kondok@faculty.chiba-u.jp}}
\email{\rm{shogo.nishino@chiba-u.jp}}
\email{\rm{sinohara@graduate.chiba-u.jp}}}
\author[2]{Shogo Nishino}
\author[2]{Toru Shinohara}

\begin{abstract}
In this paper we study the two-body bound states for gluons and ghosts in a massive Yang-Mills theory which is obtained by generalizing the ordinary massless Yang-Mills theory in a manifestly Lorentz covariant gauge.
First, we give a systematic derivation of the coupled  Bethe-Salpeter equations for gluons and ghosts by using the Cornwall-Jackiw-Tomboulis effective action of the composite operators within the framework of the path integral quantization.
Then, we obtain the numerical solutions for the Bethe-Salpeter amplitude representing the simultaneous bound states of gluons and ghosts by solving the homogeneous Bethe-Salpeter equation  in the ladder approximation.
We study how the inclusion of ghosts affects the two-gluon bound states.
Moreover, we show explicitly that the approximate solutions obtained for the gluon-gluon amplitude are consistent with the ultraviolet asymptotic freedom signaled by the negative $\beta$ function. 
\end{abstract}

\subjectindex{12.38.Aw, 21.65.Qr}

\maketitle

\section{Introduction}

As is well known, the basic ingredients relevant to  the strong interactions are quarks and gluons to be described by quantum chromodynamics (QCD).
However, we believe that they are confined and never observed in their isolated form and that only the color-singlet composite particles such as mesons, baryons and glueballs consisted of quarks and gluons can be observed in experiments. 
In particular, glueballs were predicted in the middle of the 1970s by Fritzsch and Minkowski \cite{FM75}, in contrast to mesons and baryons to be  explained by the quark model proposed by Gell-Mann in the middle of the 1960s \cite{Gell-Mann64}. Since then, the properties of glueballs have been investigated in various theoretical frameworks and also by experiments.

The equation for describing the bound states in the framework of the relativistic quantum field theory is known as the Nambu-Bethe--Salpeter equation or the Bethe--Salpeter (BS) equation \cite{Nambu50,BS51}. 
The BS equation is a self-consistent equation written in terms of the BS amplitude as the wave function of the bound state. See e.g., \cite{Nakanishi69} for reviews. 
The BS equation is specified for a given set of propagators and the interaction vertex functions describing the constituent particles. 
In the non-perturbative framework,  propagators and vertex functions can be obtained by solving the Schwinger-Dyson (SD) equation which is a self-consistent equation written in terms of propagators and vertex functions for the fundamental fields in a given field theory. 
In order to describe the bound state in the non-perturbative framework, therefore, we must take into account the SD equation and the BS equation simultaneously: 
The SD equation must be solved to give propagators and vertex functions as inputs for specifying the BS equation. Then the BS equation is to be solved in the same level of approximations as those used in solving the SD equation, as far as they cannot be solved exactly. 

The purpose of this paper is to understand glueball formation  from the underlying dynamics of the Yang-Mills theory in a way consistent with confinement and ultraviolet asymptotic freedom. 
In order to avoid complications arising from the mixing of glueballs with mesons in QCD, we restrict our consideration to the pure glue dynamics, namely, pure Yang-Mills theory. Therefore, a glueball is  supposed to be a bound state of gluons.  
In this paper, especially, we study a glueball as a two-gluon bound state for simplifying the treatment.

Already in the late 1970s, Fukuda \cite{Fukuda78} studied a two-body bound state   of gluons and ghosts as solution  of the BS equation  in the ladder approximation in the framework of the Yang-Mills theory with the covariant gauge fixing. 
He has shown the existence of the tachyon bound state due to the dominating attractive force between two gauge (gluon) particles in the color singlet  and spin singlet channel, for whatever small coupling constant. 
The gauge particles interact among themselves through a three-gluon interaction  and a contact four-gluon interaction, and interact with ghosts through the gluon-ghost interaction due to the covariant gauge fixing. 
The three-gluon interaction gives an attractive force between two gauge particles in the color singlet channel by exchanging a gauge particle (This is the analog of the attraction of opposite charge in the Abelian gauge theory).  
While the contact four-gluon interaction produces a repulsive force.  However, if the attraction due to former dominates the repulsion from the latter, the two gauge particles can form a bound state  by the net attractive force. 
This is nothing but a pairing phenomena familiar in the BCS theory of superconductivity.
Since gauge particles are massless, however, the resulting bound state should necessarily correspond to a tachyon pole.
However, his observations play the very important role  for obtaining glueballs consisted of massive gluons in this paper. 
It was also shown \cite{GM78} that such a tachyonic bound state continues to exist even in the presence of $n$ flavors of quarks and gluons acquire a mass due to vacuum rearrangement, when $n$ is smaller than or equal to a critical value $n_c$, $n \le n_c$, while all particles remain massless for $n>n_c$.

In the classical level, indeed, the Yang-Mills theory  in the four-dimensional spacetime is a massless theory with no scale. In the quantum level, however, the Yang-Mills theory should have a mass scale as suggested from the dimensional transmutation and could have the mass gap. 
Recently, a solution corresponding to a  confining gluon of the massive type was discovered as a solution of the coupled SD equation for the gluon and ghost propagators in the $SU(N)$ Yang-Mills theory with the Landau gauge fixing, which is called the decoupling solution \cite{decoupling}, in sharp contrast to the scaling solution \cite{scaling}.  The existence of the decoupling solution is supported and its confining properties are investigated by various analytical and numerical frameworks. 
However, it is not yet clear how to understand the decoupling solution as a solution to be consistent with gluon confinement and more general color confinement, see e.g., \cite{Kondo11}.

The {Brout-Englert-Higgs mechanism} is accepted as a unique way for providing the mass for gauge bosons in quantum field theories, because it is the only one established method which enables one to maintain both  {renormalizability} and {physical unitarity}. 
In this paper, we consider a specific model including a mass term for the Yang-Mills field, which is a special case of the Curci-Ferrari model \cite{CF76}. We call this model the massive Yang-Mills theory for later convenience. 
We adopt this model to perform the analytical investigation for  gluons of massive type, instead of using the decoupling solution obtained by solving the SD equation, since the decoupling solution has been  so far obtained only in a numerical way, and the analytical expression for the decoupling solution is  not yet known explicitly.

We started our analysis on massive Yang-Mills theory in the previous papers \cite{KSFNS13} without introducing the Higgs field which causes the Higgs mechanism.  
Nevertheless, we are not plagued by the unitarity violation in high energy region of the massive Yang-Mills theory without the Higgs scalar. 
This is because we regard our massive Yang-Mills theory just as a low-energy effective theory (valid below a certain cutoff $\Lambda$) of the quantum Yang-Mills theory with mass gap and confinement. 
To mimic precisely the decoupling solution, we must introduce the momentum-dependent gluon mass $m(p)$  which vanishes in the high-energy region and remains non-zero in the low-energy region. 
In this paper, we have adopted a constant mass $m$ valid up to a momentum cutoff $\Lambda$ to simplify the analyses without assuming the specific function of the momentum $p$. 
Our standpoint mentioned above has been already given   more explicitly in the second paper of \cite{KSFNS13} and is not repeated here.

It is also known that this model of massive Yang-Mills theory is still renormalizable, but loses the physical unitarity as least in the perturbation theory  \cite{KSFNS13}.
We leave the issue how to realize it as a sound model  in a quantum field theoretical framework. 
We have a hope that this issue will be resolved in a non-perturbative framework. 
One of the motivation of this paper is to obtain some information towards this direction to consider bound states in a  {massive Yang-Mills theory without the Higgs field} as a low-energy effective theory of QCD.

The issue of a tachyon bound state will be avoided by starting from the massive Yang-Mills particles instead of the massless gauge particles.  Consequently, we will obtain the massive bound states consisted of two massive gluons described by the massive Yang-Mills theory.  
In obtaining such a solution, especially, we pay special attention to the aspect how our results for the bound state is consistent with the ultraviolet asymptotic freedom  \cite{Fukamachi15}. 
This is a feature that has not been discussed in the preceding works \cite{MS13,SAFKS15}.
In their works, the precise determinations of the glueball spectrum have been the main purpose of  investigations. 
Such an issue is not the aim of the present paper, although it is a nice goal of future investigations to be achieved by developing our method. 

We give a systematic derivation of the coupled  BS equation for gluon and ghost by starting from the the Cornwall-Jackiw-Toumboulis effective (CJT) action for the composite operator with the bilocal source term. 
We show that the bound state obtained as a solution of the homogeneous BS equation for the gluon-gluon BS amplitude is consistent with the ultraviolet asymptotic freedom. 

This paper is organized as follows.
In sec. II, we introduce the massive Yang-Mills theory from which we start the study of the bound state.
In section III, we give the CJT effective action for the bilocal composite operator, which gives a systematic derivation of the BS equation. 
In sec. IV, we derive the BS equation by differentiating the CJT effective action with respect to the full Green functions. 
In sec. V, we write down the coupled BS equation for gluon and ghost at vanishing total momentum $P=0$ in the massless case, and obtain the numerical solutions. This is a preliminary study of obtaining the bound state. 
In sec. VI, we write down the homogeneous BS equation at the general total momentum $P$ in the massive case, but restricting to the gluon-gluon BS amplitude alone. 
In sec. VII, 
we solve the BS equation for the gluon-gluon BS amplitude at the general total momentum $P$ in the massive case. 
The final section is devoted to conclusion and discussion. 
Some technical materials are collected in the Appendices.

\section{The massive Yang-Mills theory}

In order to discuss the bound state, we adopt the massive $SU(N)$ Yang-Mills theory without the Higgs field, which is defined to be the usual massless $SU(N)$ Yang-Mills theory in the most general Lorenz gauge \cite{Baulieu85} plus the ``mass term'' $\mathscr{L}_m$ formulated in a manifestly Lorentz covariant way. 
The total Lagrangian density $\mathscr{L}^{\rm{tot}}_{m\rm{YM}}$ is written in terms of the Yang-Mills field $\mathscr{A}_\mu$, the  Faddeev-Popov ghost field $\mathscr{C}$, the antighost field $\bar{\mathscr{C}}$ and the Nakanishi-Lautrup auxiliary field $\mathscr{N}$, if we use the terminology adopted in the usual massless Yang-Mills theory:
\begin{subequations}
	\begin{align}
		\mathscr{L}^{\rm{tot}}_{m\rm{YM}} =& \mathscr{L}_{\rm{YM}} + \mathscr{L}_{\rm{GF+FP}} + \mathscr{L}_{m} , 
\\
		\mathscr{L}_{\rm{YM}}  =& - \frac{1}{4} \mathscr{F}_{\mu \nu} \cdot \mathscr{F}^{\mu \nu} , 
   \\
		\mathscr{L}_{\rm{GF+FP}}  
		 =& \mathscr{N} \cdot \partial^{\mu} \mathscr{A}_{\mu} + i \bar{\mathscr{C}} \cdot \partial^{\mu} \mathscr{D}_{\mu}[\mathscr{A}] \mathscr{C}
+ \frac{\beta}{4} ( \bar{\mathscr{N}} \cdot \bar{\mathscr{N}} + \mathscr{N} \cdot \mathscr{N}) 
		+ \frac{\alpha}{2} \mathscr{N} \cdot \mathscr{N}
\nonumber\\
=& \frac{\alpha}{2} \mathscr{N} \cdot \mathscr{N}  + \frac{\beta}{2} \mathscr{N} \cdot \mathscr{N} 
         + \mathscr{N} \cdot \partial^{\mu} \mathscr{A}_{\mu} 
		- \frac{\beta}{2} g \mathscr{N} \cdot (i \bar{\mathscr{C}} \times \mathscr{C}) 
  \nonumber\\
		& + i \bar{\mathscr{C}} \cdot \partial^{\mu} \mathscr{D}_{\mu}[\mathscr{A}] \mathscr{C}
		+ \frac{\beta}{4} g^2 (i \bar{\mathscr{C}} \times \mathscr{C}) \cdot (i \bar{\mathscr{C}} \times \mathscr{C}) 
, 
   \\
		\mathscr{L}_{m}  =& \frac{1}{2} m^2 \mathscr{A}_{\mu} \cdot \mathscr{A}^{\mu} + \beta m^2 i \bar{\mathscr{C}} \cdot \mathscr{C} , 
	\end{align}
\end{subequations}
where  
$\mathscr{D}_{\mu}$ is the covariant derivative defined by
	\begin{align}
 \mathscr{D}_{\mu}[\mathscr{A}] \omega(x) 
:= \partial_{\mu} \omega(x) + g \mathscr{A}_\mu(x) \times \omega(x) ,
	\end{align}
and 
$\bar{\mathscr{N}}$ is defined by
\begin{equation}
\bar{\mathscr{N}} :=-\mathscr{N}+gi\bar{\mathscr{C}} \times \mathscr{C} .
\end{equation}
Here we have adopted the notation:
\begin{align}
\mathscr{X} \cdot \mathscr{Y}
 :=& \mathscr{X}^a \mathscr{Y}^a ,
\quad 
(\mathscr{X} \times \mathscr{Y})^c  
 := f_{abc} \mathscr{X}^a \mathscr{Y}^b  ,
\end{align}
where $f_{abc}$ is the structure constants of the Lie algebra of the $SU(N)$ group. 
Here $\alpha$ and $\beta$ are parameters which are identified with the gauge-fixing parameters in the $m \rightarrow 0$ limit.
The $\alpha=0$ case is the Curci-Ferrari model \cite{CF76} with the coupling constant $g$, the mass parameter $m$ and the parameter $\beta$.
In the Abelian limit with  vanishing structure constants $f_{abc}=0$, the Faddeev-Popov ghosts decouple and the Curci-Ferrari model reduces to the Nakanishi model \cite{Nakanishi72}.

In what follows, we put $\beta=0$ to simulate the decoupling solution representing massive gluon and massless ghost. 
By eliminating the Nakanishi-Lautrup field $\mathscr{N}$, $\mathscr{L}^{\rm{tot}}_{m\rm{YM}}$ reads
\begin{align}
\mathscr{L}^{\rm{tot}}_{m\rm{YM}} 
=&-\frac{1}{4}(\partial _\mu \mathscr{A}_{\nu} -\partial _\nu \mathscr{A}_{\mu}) \cdot  (\partial ^\mu \mathscr{A}^{\nu} -\partial ^\nu \mathscr{A}^{\mu})  
- \frac{1}{2\alpha} \left( \partial ^\mu \mathscr{A}_{\mu} \right) ^2
+ i\bar{\mathscr{C}}  \cdot \Box \mathscr{C} 
\notag \\
&
-\frac{1}{2}g\left( \partial _\mu \mathscr{A}_{\nu} -\partial _\nu \mathscr{A}_{\mu} \right)  \cdot \left( \mathscr{A}^{\mu} \times \mathscr{A}^{\nu} \right) 
-\frac{1}{4}g^2\left( \mathscr{A}_{\mu} \times \mathscr{A}_{\nu} \right) ^2
+g i\bar{\mathscr{C}}  \cdot \partial ^\mu \left( \mathscr{A}_{\mu} \times \mathscr{C} \right) 
\notag \\
&+\frac{1}{2}m^2 \mathscr{A}_{\mu}  \cdot \mathscr{A}^{\mu} .
\label{mYM}
\end{align}
By starting from the massive Yang-Mills theory given by the Lagrangian density (\ref{mYM}), we derive the BS equation which includes simultaneously  gluon and ghost on equal footing. 
This can be done using the CJT effective action for the composite operators, as explicitly shown in \cite{Fukamachi15}.
See \cite{KSFNS13} for more details on this model.

\section{Cornwall--Jackiw--Tomboulis effective action}


For the Yang--Mills theory, we introduce the \textbf{local sources} $J^\mu _a(x)$, $\bar{\eta}_a(x)$, and $\eta _a(x)$ for the \textbf{gluon field} $\mathscr{A}_\mu ^a(x)$, the \textbf{ghost field} $\mathscr{C}^a(x)$, and the \textbf{antighost field} $\mathscr{\bar{C}}^a(x)$ respectively.
Moreover, we introduce the \textbf{bilocal sources} $I_{ab}^{\mu \nu}(x,y)$ and $\theta _{ab}(x,y)$ for the bilocal composite operators $\mathscr{A}_\mu^a(x) \mathscr{A}_\nu^b(y)$ and $\mathscr{\bar{C}}^a(x) \mathscr{C}^b(y)$ respectively,
where the bilocal source $I$ is symmetric $I^{ab}_{\mu \nu}(x,y)=I^{ba}_{\nu \mu}(y,x)$ and $\theta ^{ab}(x,y)$ is a general matrix. 
Then the generating functional $W[J,\bar{\eta},\eta ,I,\theta ]$ is defined by 
\begin{align}
& Z[J,\bar{\eta},\eta ,I,\theta ]
= e^{iW[J,\bar{\eta},\eta ,I,\theta ]} 
\notag \\
=&\int \mathcal{D}\Phi \exp \Bigg\{ iS^{\rm{tot}}_{m\rm{YM}}[\mathscr{A},\mathscr{\bar{C}},\mathscr{C}]+i\int_{x} \left[ J^\mu _a(x)\mathscr{A}_\mu ^a(x)+\bar{\eta}_a(x)\mathscr{C}^a(x)+ \mathscr{\bar{C}}^a(x)\eta _a(x) \right] 
\notag \\
&\hspace{3cm}
+i \int_{x} \int_{y} \left[ \frac{1}{2}\mathscr{A}^a_\mu (x)I_{ab}^{\mu \nu}(x,y)\mathscr{A}^b_\nu (y)+ \mathscr{\bar{C}}^a(x)\theta _{ab}(x,y) \mathscr{C}^b(y)\right] \Bigg\} ,
\end{align}
where we have introduced the abbreviated notation: 
$\int_{x}:=\int d^Dx$.
For the local source, we find that the (left) derivatives of $W$ with respect to the  local sources are given by
\begin{subequations}
\begin{align}
\frac{\delta W}{\delta J_a^\mu (x)}&=\left< \mathscr{A}^a_\mu (x)\right> \equiv \bm{A}_\mu ^a(x) ,
\\
\quad
\frac{\delta W}{\delta \bar{\eta}_a(x')} &= \left< \mathscr{C}^a(x')\right> \equiv \bm{C}^a(x') ,
\quad
\frac{\delta W}{\delta \eta _a(x')} =-\left< \mathscr{\bar{C}}^a(x')\right> \equiv -\bar{\bm{C}}^a(x') ,
\\
\frac{\delta ^2W}{\delta J^\nu _b(y)\delta J_a^\mu (x)}&=i\left< \mathscr{A}^a_\mu (x); \mathscr{A}^b_\nu (y)\right> \equiv iD^{ab}_{\mu \nu}(x,y) ,
\\
\frac{\delta ^2W}{\delta \eta _b(y')\delta \bar{\eta}_a(x')} &= i\left< \mathscr{C}^a(x');\mathscr{\bar{C}}^b(y')\right> \equiv i\Delta ^{ab}(x',y') ,
\\
\frac{\delta ^2W}{\delta \bar{\eta}_b(y')\delta \eta _a(x')} &=i\left< \mathscr{\bar{C}}^a(x');\mathscr{C}^b(y)\right> \equiv i\tilde{\Delta}^{ab}(x',y') ,
\end{align}
\end{subequations}
where $D$ and $\tilde{\Delta}$ are the full gluon and ghost-antighost propagators.

For the bilocal source, we find that the  derivatives of $W$ with respect to the bilocal sources are given by
\begin{subequations}
\begin{align}
\frac{\delta W}{\delta I_{ab}^{\mu \nu}(x,y)}&
=\left< \mathscr{A}_\mu ^a(x) \mathscr{A}_\nu ^b(y)\right> =\left< \mathscr{A}_\mu ^a(x); \mathscr{A}_\nu ^b(y)\right> +\left< \mathscr{A}_\mu ^a(x)\right> \left< \mathscr{A}_\nu ^b(y)\right> \notag \\
&=D_{\mu \nu}^{ab}(x,y)+\bm{A}_\mu ^a(x)\bm{A}_\nu ^b(y) ,
\\
\frac{\delta W}{\delta \theta _{ab}(x',y')}&
=\left< \mathscr{\bar{C}}^a(x')\mathscr{C}^b(y')\right> =\left< \mathscr{\bar{C}}^a(x');\mathscr{C}^b(y')\right> +\left< \mathscr{\bar{C}}^a(x')\right> \left< \mathscr{C}^b(y')\right> \notag \\
&=\tilde{\Delta} ^{ab}(x',y')+\bar{\bm{C}}^a(x')\bm{C}^b(y') .
\end{align}
\end{subequations}

We define the CJT effective action \cite{CJT74} of the Yang-Mills theory by the Legendre transform:
\begin{align}
\Gamma [ \bm{A},\bar{\bm{C}},\bm{C},D,\tilde{\Delta} ]=&W[J,\bar{\eta},\eta ,I,\theta ] - \int_{x}  \left( J_a^\mu (x)\bm{A}^a_\mu (x)+\bar{\eta}_a(x)\bm{C}^a(x)+\bar{\bm{C}}^a(x)\eta _a(x) \right) 
\notag \\
&-\frac{1}{2} \int_{x} \int_{y} \left( \bm{A}^a_\mu (x)I_{ab}^{\mu \nu}(x,y)\bm{A}^b_\nu (y)+I_{ab}^{\mu \nu}(x,y)D^{ab}_{\mu \nu}(x,y)\right) \notag \\
&- \int_{x'} \int_{y'} \left( \bar{\bm{C}}^a(x')\theta _{ab}(x',y')\bm{C}^b(y')+\theta _{ab}(x',y')\tilde{\Delta}^{ab} (x',y')\right) .
\label{effective_action}
\end{align}
Then we find the derivatives of $\Gamma$ with respect to the field expectation,  
\begin{subequations}
\begin{align}
\frac{\delta \Gamma}{\delta \bm{A}_\mu ^a(x)}&=-J^\mu _a(x)- \int_{y} I_{ab}^{\mu \nu}(x,y)\bm{A}_\nu ^b(y) ,
\\
\frac{\delta \Gamma}{\delta \bar{\bm{C}}^a(x')}&= - \eta _a(x')- \int_{y'} \theta _{ab}(x',y')\bm{C}^b(y') , 
\quad
\frac{\delta \Gamma}{\delta \bm{C}^b(y')} = \bar{\eta}_b(y')+ \int_{y'} \bar{\bm{C}}^a(x')\theta _{ab}(x',y') ,
\end{align}
and the derivatives of  $\Gamma$ with respect to the full propagators, 
\begin{align}
\frac{\delta \Gamma}{\delta D_{\mu \nu}^{ab}(x,y)}&=-I^{\mu \nu}_{ab}(x,y) ,
\quad
\frac{\delta \Gamma}{\delta \tilde{\Delta} ^{ab}(x',y')} =-\theta _{ab}(x',y') .
\end{align}
\end{subequations}

It is shown that the CJT effective action is rewritten as
\begin{align}
\Gamma [\bm{A},\bm{C},\bar{\bm{C}},D,\tilde{\Delta}]
=& S^{\rm{tot}}_{m\rm{YM}}[\bm{A},\bm{C},\bar{\bm{C}}] + \frac{i}{2}{\text{Tr}} \left[ \mathscr{D}^{-1}_{AA}D\right] - i{\text{Tr}}\left[ \bar{\mathscr{D}}^{-1}_{C\bar{C}}\tilde{\Delta}\right] 
\notag \\
&+ \frac{i}{2}{\text{Tr}}\ln D^{-1} + i{\text{Tr}}\ln \tilde{\Delta}^{-1} 
+ \Gamma _2[\bm{A},\bm{C},\bar{\bm{C}},D,\tilde{\Delta}] ,
\end{align}
where $\Gamma_2$ is the two particle irreducible (2PI) contributions, 
$D$ and $\tilde{\Delta}$ are respectively the full gluon and ghost propagators, $\mathscr{D}_{AA}$ and $\bar{\mathscr{D}}_{C\bar{C}}$ are respectively the tree-level gluon and ghost propagators defined by 
\begin{align}
\frac{\delta ^2 \bm{S}}{\delta \bm{A}_\nu ^b(y)\delta \bm{A}_\mu ^a(x)}&
\equiv  (i\mathscr{D}^{-1}_{AA})_{\mu\nu}^{ab}(x,y) ,
\quad
\frac{\delta ^2 \bm{S}}{\delta \bar{\bm{C}}^a(x) \delta \bm{C}^b(y)} 
\equiv (i\bar{\mathscr{D}}^{-1}_{C\bar{C}})^{ab}(x,y) .
\end{align}

\begin{figure}[t]
\centering
\includegraphics[scale=0.54]{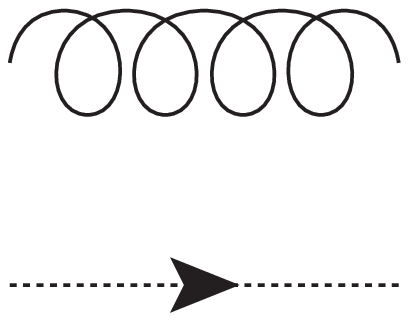}
\quad\quad
\includegraphics[scale=0.37]{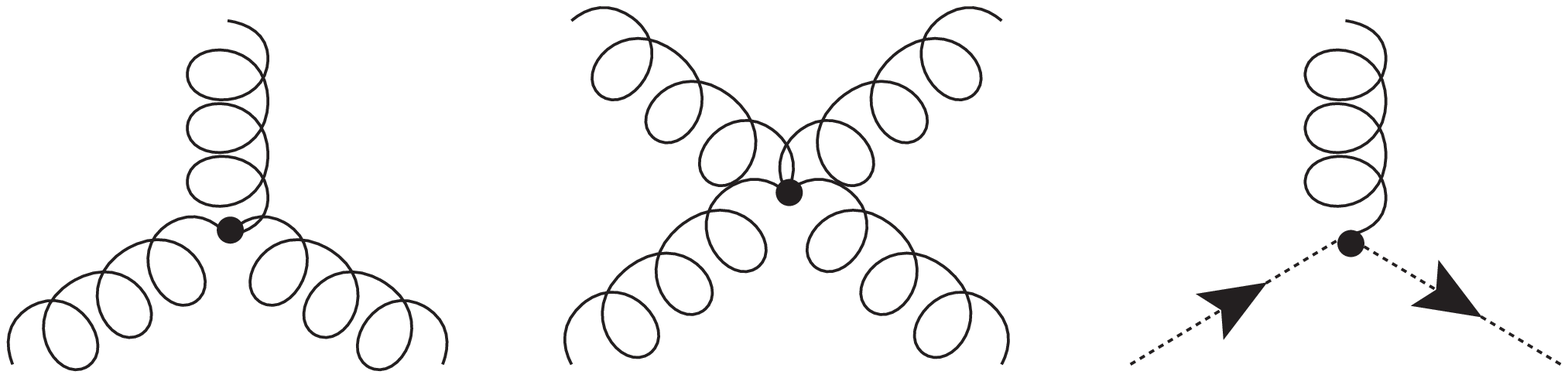}
\caption{
Bare gluon and ghost propagators;  
the bare vertex functions: 3-gluon, 4-gluon, gluon-ghost-antighost vertices  in the Yang-Mills theory. 
}
\label{fig:propagator-vertex}
\end{figure}

\begin{figure}[t]
\centering
\includegraphics[scale=0.54]{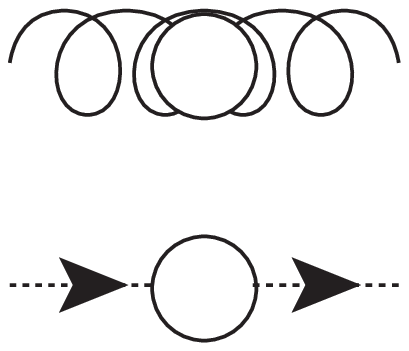}
\quad\quad
\includegraphics[scale=0.37]{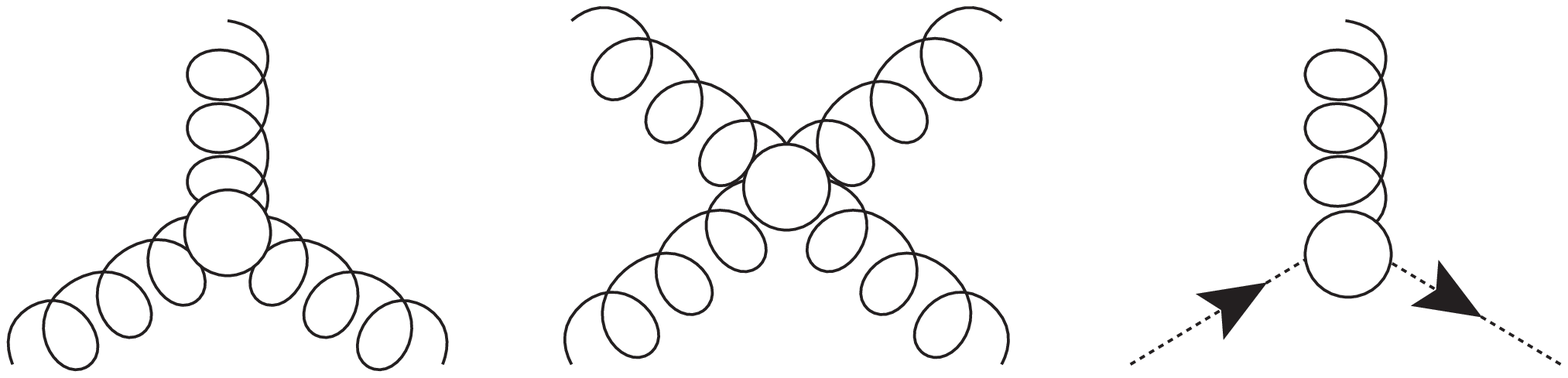}
\caption{
Full gluon and ghost propagators;  
the full vertex functions: 3-gluon, 4-gluon, gluon-ghost-antighost vertices  in the Yang-Mills theory. 
}
\label{fig:propagator-vertex-full}
\end{figure}

\begin{figure}[t]
\includegraphics[scale=0.42]{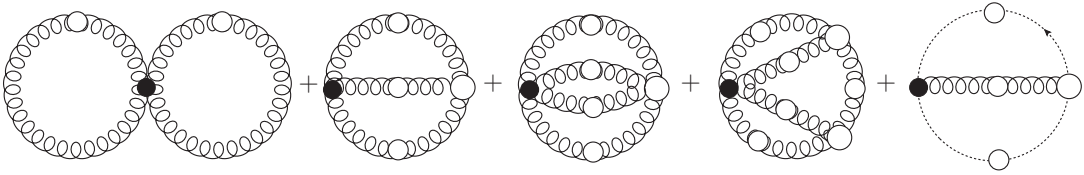}
\caption{
2PI vacuum diagrams in the Yang-Mills theory.
}
\label{fig:vacuum-diagram}
\end{figure}

The original Yang-Mills theory is characterized by the bare gluon and ghost propagators, and the bare one particle irreducible (1PI) vertex functions: 3-gluon, 4-gluon, and gluon-ghost-antighost vertices given in Fig.~\ref{fig:propagator-vertex}. 
The full gluon and ghost propagators, and the full  1PI vertex functions: 3-gluon, 4-gluon, and gluon-ghost-antighost vertices are given in Fig.~\ref{fig:propagator-vertex-full}. 
It is shown that the 2PI vacuum diagrams in the Yang-Mills theory are written as Fig.~\ref{fig:vacuum-diagram}.


\section{A systematic derivation of the Bethe--Salpeter equation}

\subsection{Derivation of the Schwinger--Dyson equation}

\begin{figure}[h]
\centering
\includegraphics[scale=0.42]{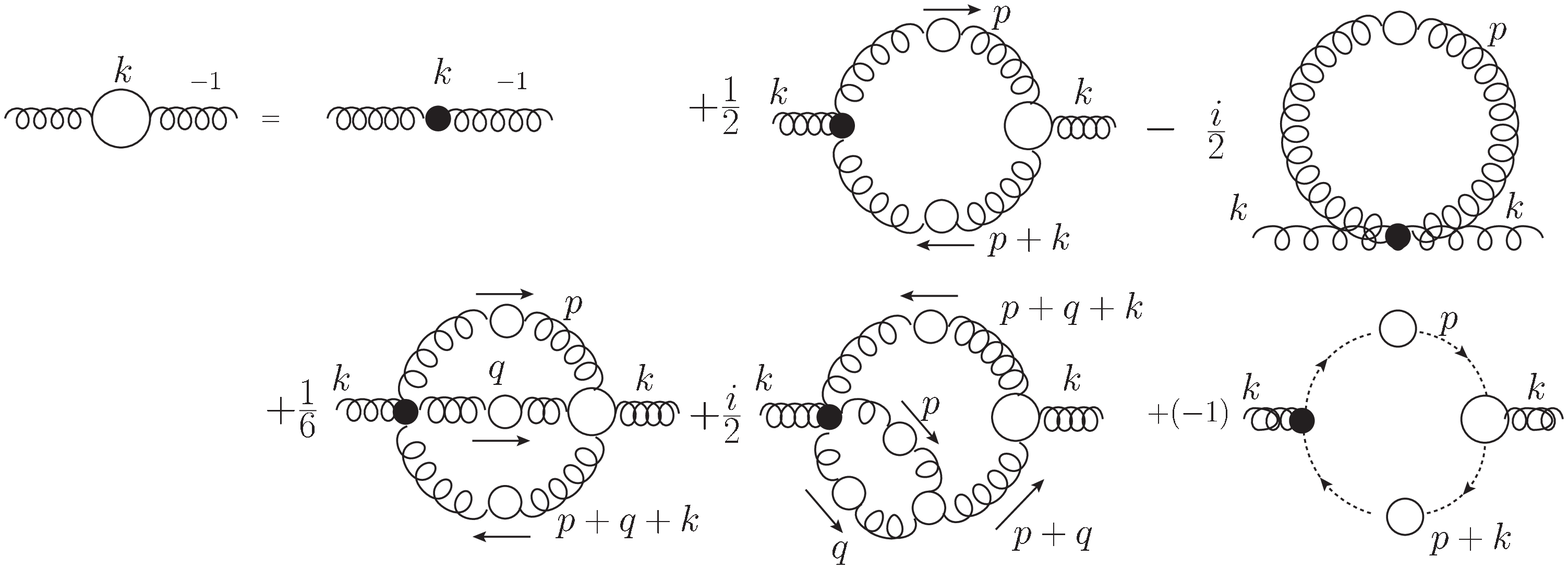}
\caption{
Gluon-SD equation. 
}
\label{fig:gluon-SD}

\includegraphics[scale=0.40]{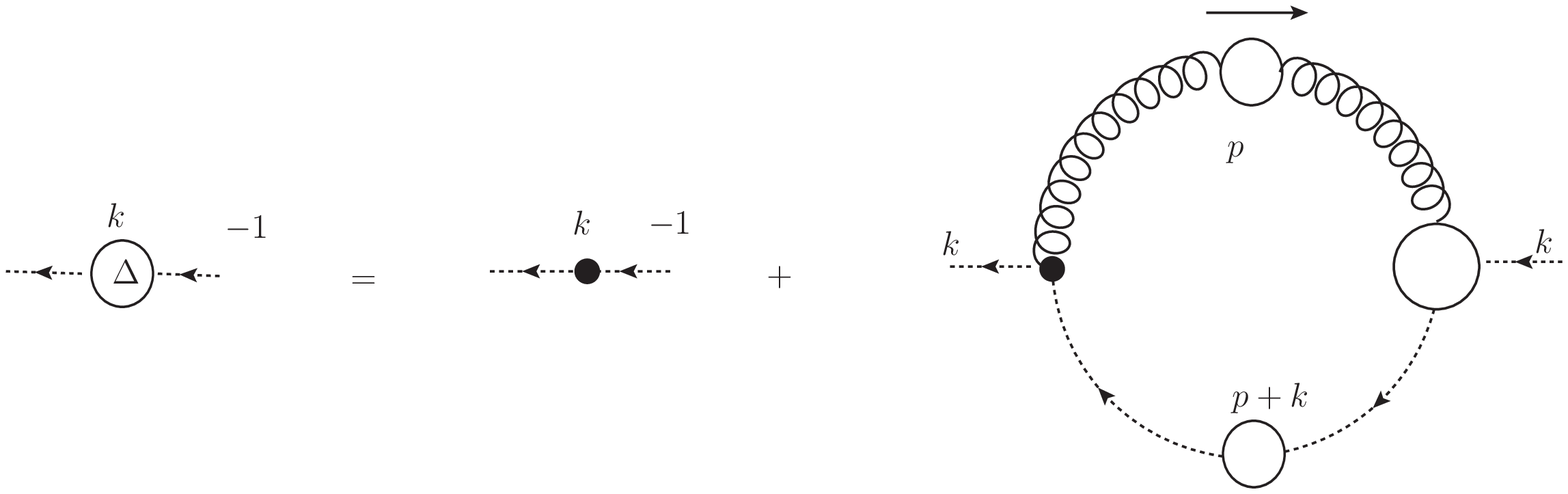}

\caption{
Ghost-SD equation. 
}
\label{fig:ghost-SD}
\end{figure}

By taking the first derivative of  the CJT effective action $\Gamma$ with respect to the full gluon and ghost propagators $D$, we obtain the SD equations for the gluon propagator $D$ and ghost propagator $\delta \tilde{\Delta}$ respectively (after  the sources are setting to be zero $I, \theta \to 0$): 
\begin{subequations}
\begin{align}
 -I = \frac{\delta \Gamma}{\delta D} = -iD^{-1} + i\mathscr{D}_{AA}^{-1}+ \frac{\delta \Gamma_2}{\delta D} ,
\\
    -\theta = \frac{\delta \Gamma}{\delta \tilde{\Delta}(x,y)} = -i\tilde{\Delta}^{-1} + i\tilde{\mathscr{D}}_{C\bar{C}}^{-1}+ \frac{\delta \Gamma_2}{\delta \tilde{\Delta}} .
\end{align}
\end{subequations}
Fig.~\ref{fig:gluon-SD} is the gluon SD equation and Fig.~\ref{fig:ghost-SD} is the ghost SD equation. 

\subsection{The equation for the scattering Green function}

The BS equation is obtained by taking the first derivative of the SD equation (i.e., the second derivative of the CJT effective action) with respect to the full propagator.
For the general field $\phi$, the \textbf{scattering Green function} $D_4$ with four external lines is defined by  the second derivative of $W$ with respect to the bilocal source $I$:
\begin{align}
D_4 (x^\prime,y^\prime;x,y)
=& \left< 0|\phi (x)\phi  (y);\phi (x^\prime)\phi (y^\prime)|0\right>  
= \frac{1}{i} \frac{\delta ^2 W}{\delta I (x^\prime,y^\prime)\delta I (x,y)} ,
\end{align}
while the inverse $D_4^{-1}$ of the {scattering Green function} $D_4$ is given by  the second derivative of $\Gamma$ with respect to the full propagator $D$:
\begin{align}
 D_4^{-1} := \frac{\delta^2 \Gamma}{\delta D\delta D} .
\end{align}
Therefore, the first derivative of the SD equation yields the equation for the inverse $D_4^{-1}$ of the  {scattering Green function} $D_4$:
\begin{align}
 D_4^{-1} = D_4^{(0)}{}^{-1}  + K , 
\quad D_4^{(0)}{}^{-1} := \frac{\delta^2 [-iD^{-1}]}{\delta D\delta D} , 
\quad K := \frac{\delta^2 \Gamma_2}{\delta D\delta D} ,
\end{align}
where $D_4^{(0)}$ is the tree part of the scattering Green function $D_4$ and $K$ denotes the kernel. 
The diagrammatic representation is given as 
\begin{equation}
\includegraphics[scale=0.36]{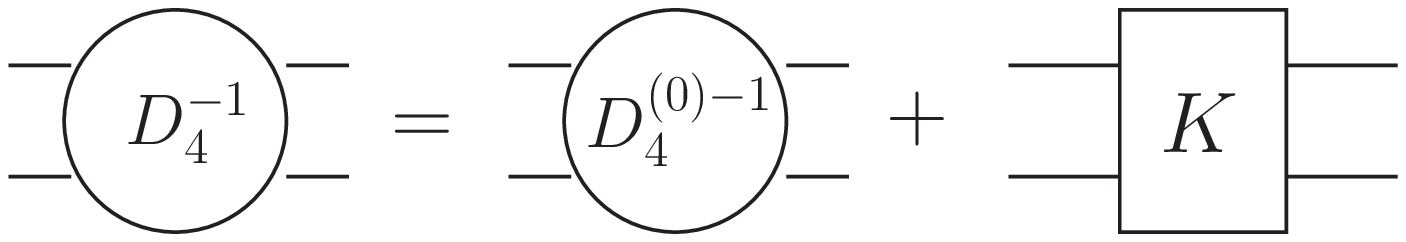}
\end{equation}
We make use of the identities:
\begin{equation}
\includegraphics[scale=0.30]{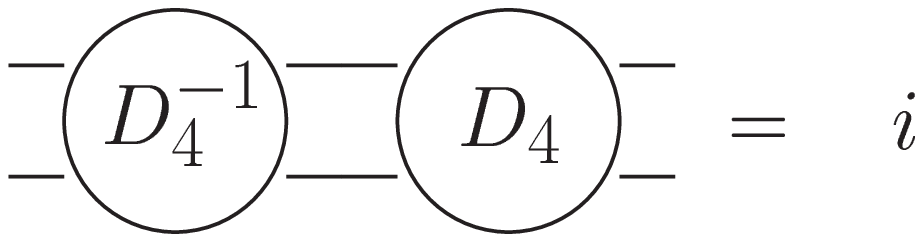}\hspace{1.5cm} {\rm{and}} \hspace{1.5cm} \includegraphics[scale=0.30]{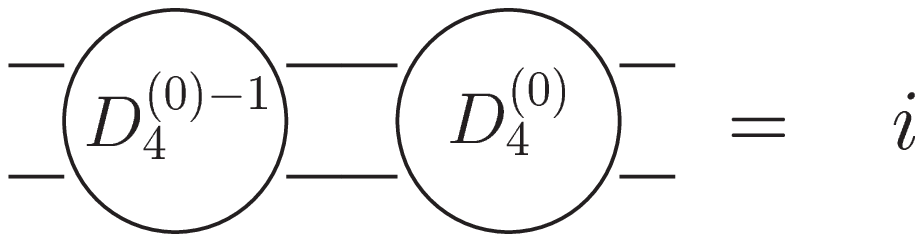}
\end{equation}
to rewrite the equation in terms of $D_4$, not the inverse $D_4^{-1}$. 
We multiply $D_4$ from the right to obtain 
\begin{equation}
\includegraphics[scale=0.30]{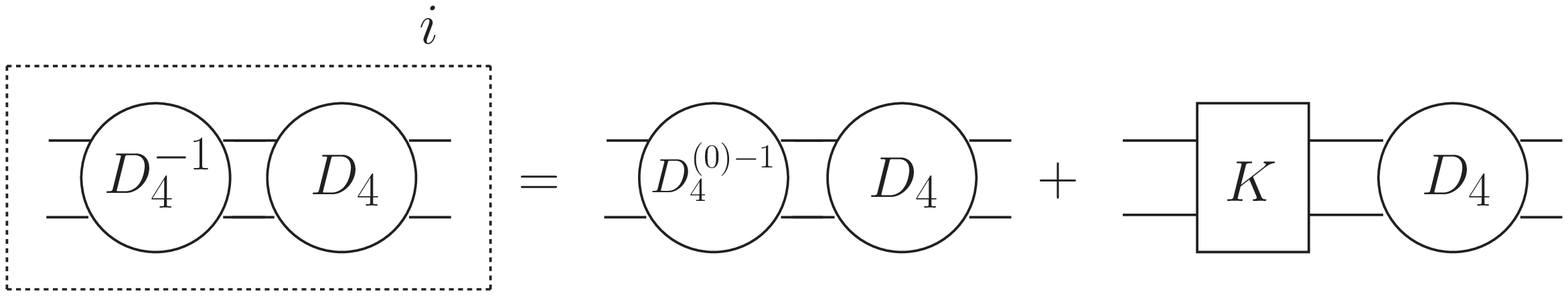}
\end{equation}
and then multiply $D_4^{(0)}$ from the left to obtain

\begin{equation}
\includegraphics[scale=0.30]{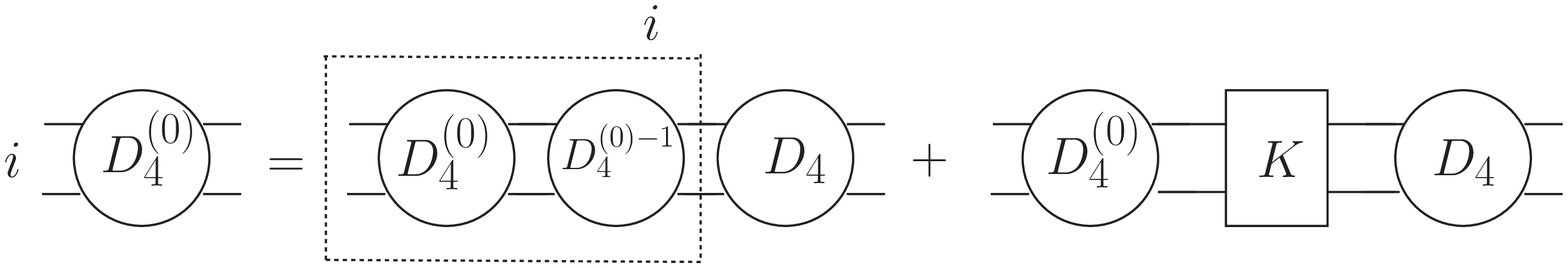}
\end{equation}

Thus we obtain the self-consistent equation for the scattering Green function $D_4$ after multiplying $i$:

\begin{equation}
\includegraphics[scale=0.36]{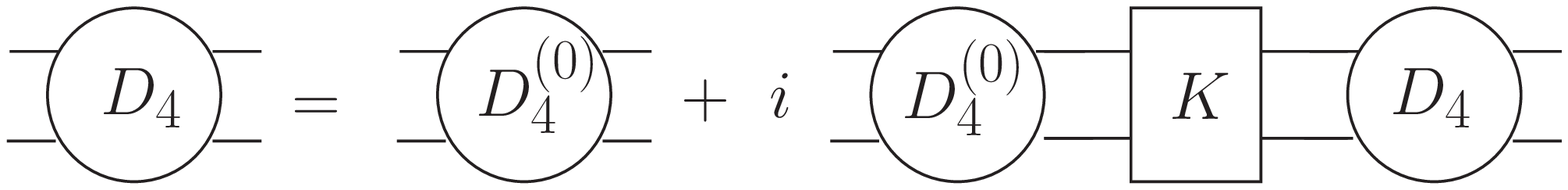}
\end{equation}

\begin{figure}[t]
\includegraphics[scale=0.44]{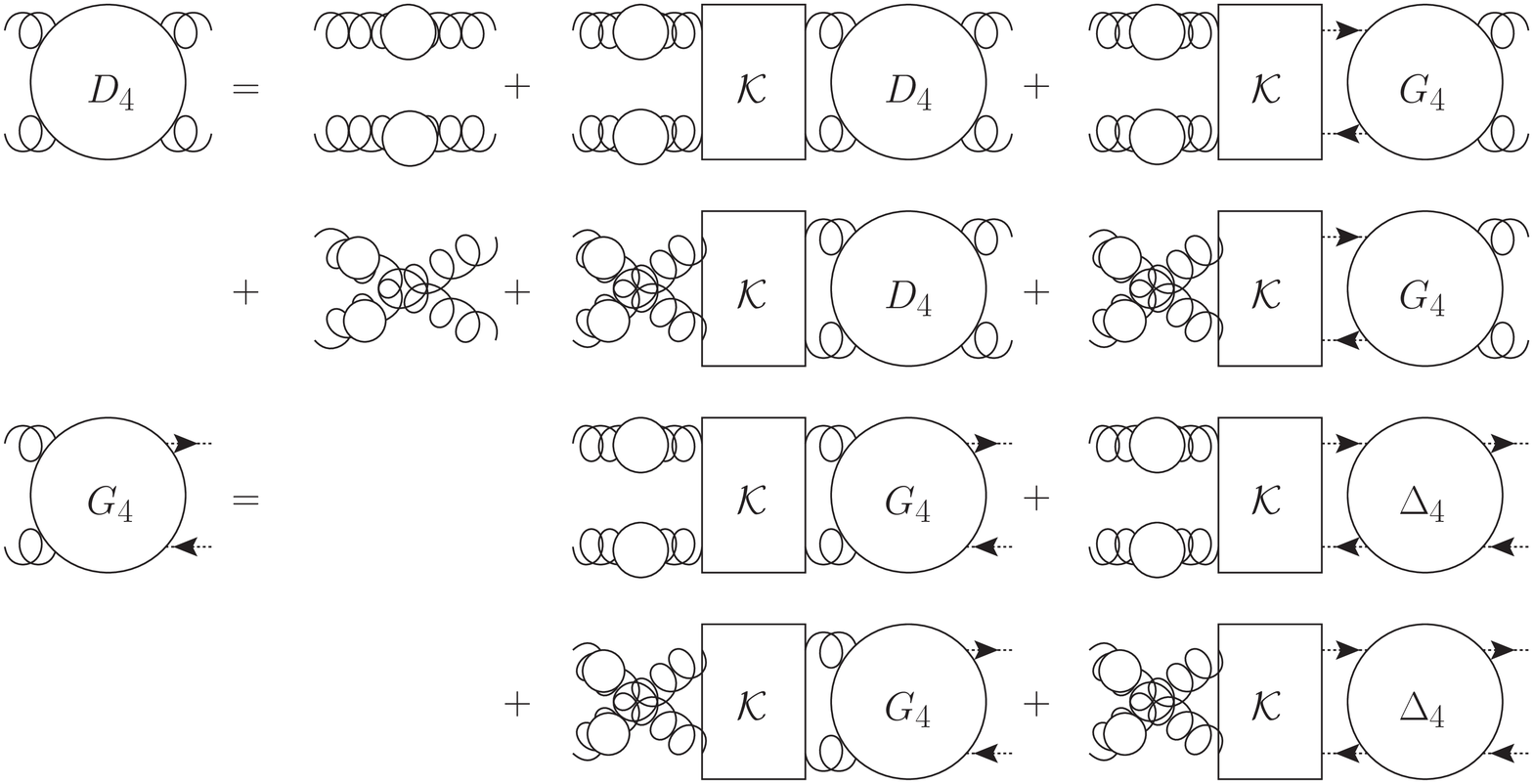}
\includegraphics[scale=0.44]{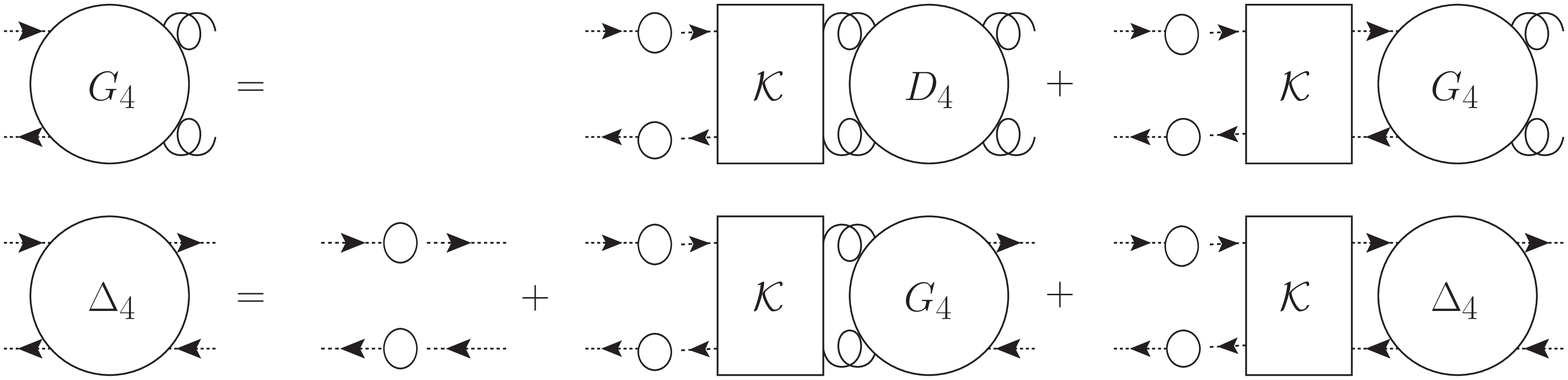}
\caption{
The  BS equation for 
gluon-gluon $D$, 
gluon-ghost $G$,
ghost-gluon $G$,
and 
ghost-ghost $\Delta$ scattering Green functions. 
}
\label{fig:mYM-BS4}
\end{figure}

For the Yang-Mills theory, we obtain a set of the BS equations for the four BS amplitude: gluon-gluon,  gluon-ghost, ghost-gluon, and ghost-ghost scattering Green function  
 as given in Fig.~\ref{fig:mYM-BS4}. 

%
%
%
%
%
%

\subsection{The homogeneous BS equation for the (amputated) BS amplitude}

We can use the completeness relation:
\begin{equation}
 \bm{1}=|0\big> \big< 0|+\sum _c \int \frac{d^d\bm{P}_c}{(2\pi )^d}\frac{1}{2E_{P_c}}|P_c\big> \big< P_c| +\cdots ,
\end{equation}
where the sum $\sum _c$ runs over all the bound states except for the continuum spectrum.
The {scattering Green function} $D_4$ for particles described by the field $\phi_a$ is defined by  
\begin{align}
D_4{}^{a'b';ab}(x',y';x,y)
=& \left< 0|\phi^a (x)\phi^b (y);\phi^{a'}(x')\phi^{b'}(y')|0\right>  
= \frac{1}{i}  \frac{\delta ^2W}{\delta I^{a'b'}(x',y')\delta I^{ab}(x,y)} .
\end{align}
This is decomposed into the product of the  \textbf{BS amplitude} $\chi$:
\begin{align}
D_4{}^{a'b';ab}(x',y';x,y)
=  &\sum _c\int \frac{d^dP_c}{(2\pi )^d}\frac{1}{2E_{P_c}}\chi ^{ab}_{P}(x,y)\bar{\chi}^{a'b'}_{P}(x',y') + ... ,
\end{align}
where the BS amplitude $\chi$ is defined by
\begin{equation}
\chi ^{ab}_{P}(x,y)=\left< 0|\phi^a (x)\phi^b (y)|P_c\right>    .
\end{equation}
The momentum space representation of $D_4$ has  the pole: 
\begin{align}
D_4{}^{a'b';ab} (q,p;P) =& \sum _c \frac{i}{(2\pi )^D}\frac{\chi ^{ab}_{P }(p)\bar{\chi}^{a'b'}_{P }(q)}{P^2-M^2} + ...,
\end{align}
the  product $\bar{\chi} \chi$ of the BS amplitude corresponds to the residue at the pole where the squared momentum $P^2$ coincides with the bound state mass $M$ squared: $P^2=M^2=P_c^2$.

\begin{equation}
\includegraphics[scale=0.3]{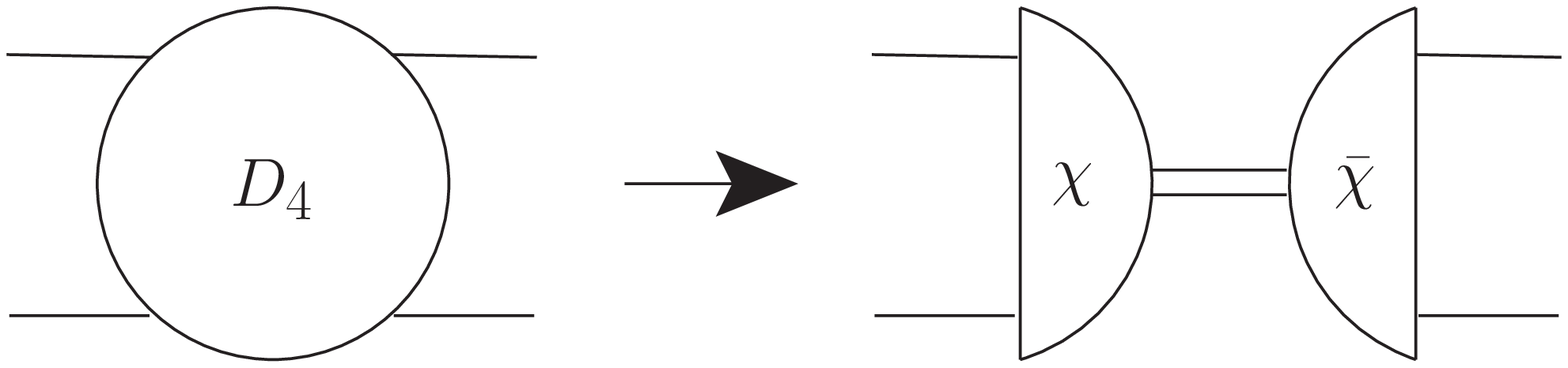}
\end{equation}


From the equation for the scattering Green function $D_4$,
\begin{equation}
\includegraphics[scale=0.35]{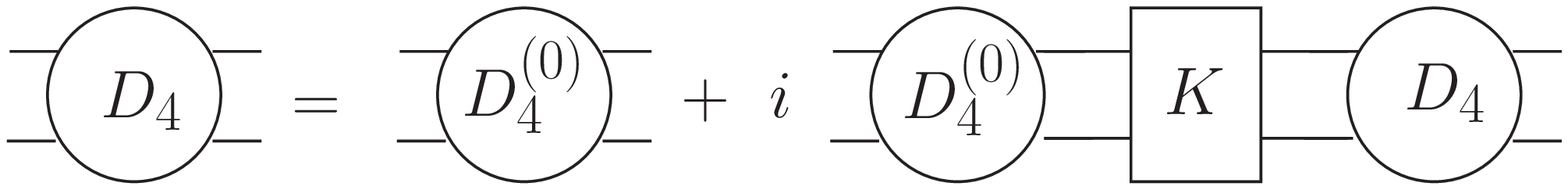}
\end{equation}
we can extract the product $\bar{\chi} \chi$ of the BS amplitude by identifying it with the pole residue  at the pole corresponding to the bound state, since $D_4^{(0)}$ does not have the pole:
\begin{equation}
\includegraphics[scale=0.3]{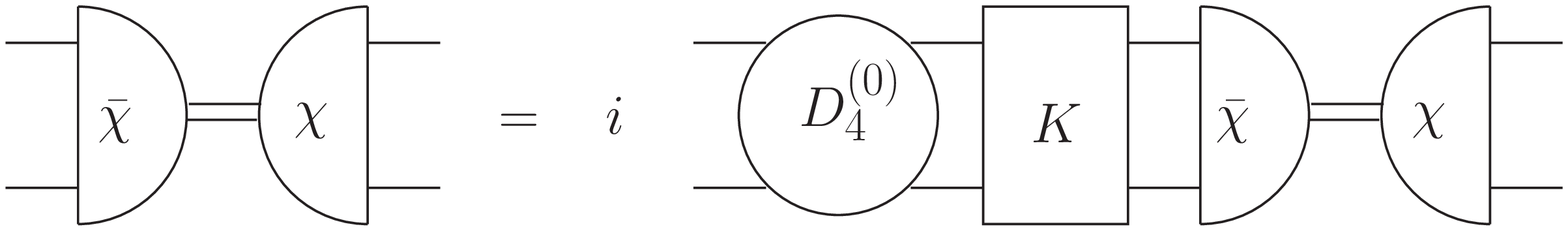}
\end{equation}
By multiplying $D_4^{(0)}{}^{-1}$ from the left,
\begin{equation}
\includegraphics[scale=0.3]{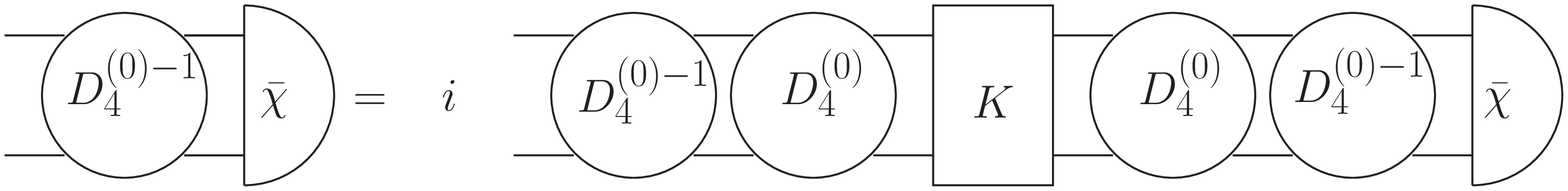}
\end{equation}
we finally obtain the \textbf{homogeneous BS equation} for the \textbf{amputated BS amplitude} $D^{(0)}{}^{-1}\bar\chi$:
\begin{equation}
\includegraphics[scale=0.3]{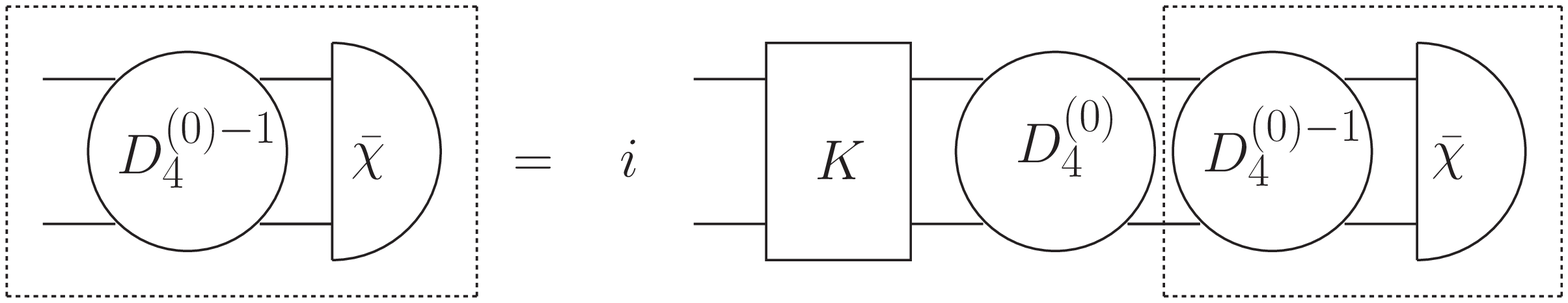}
\end{equation}


\rule[0cm]{0cm}{0cm}

For the Yang-Mills theory, the {scattering Green function} $D_4$  for the gluon field is defined by  
\begin{align}
D_4{}^{a'b';ab}_{\mu '\nu ';\mu \nu}(x',y';x,y)
=& \left< 0|\mathscr{A}^a_\mu (x)\mathscr{A}^b_\nu (y);\mathscr{A}^{a'}_{\mu '}(x')\mathscr{A}^{b'}_{\nu '}(y')|0\right>  
= \frac{1}{i}  \frac{\delta ^2W}{\delta I_{\mu '\nu '}^{a'b'}(x',y')\delta I_{\mu \nu}^{ab}(x,y)} .
\end{align}
This is decomposed into the product of the  \textbf{BS amplitude} $\chi$:
\begin{align}
D_4{}^{a'b';ab}_{\mu '\nu ';\mu \nu}(x',y';x,y)
=  &\sum _c\int \frac{d^dP_c}{(2\pi )^d}\frac{1}{2E_{P_c}}\chi ^{ab}_{P;\mu \nu}(x,y)\bar{\chi}^{a'b'}_{P;\mu '\nu '}(x',y') + ... .
\end{align}
In the momentum space representation, we have
\begin{align}
D_4{}^{a'b';ab}_{\mu '\nu ';\mu \nu}(q,p;P) =& \sum _c \frac{i}{(2\pi )^D}\frac{\chi ^{ab}_{P;\mu \nu}(p)\bar{\chi}^{a'b'}_{P;\mu '\nu '}(q)}{P^2-M^2} + ...,
\end{align}
We introduce the BS amplitudes $\chi ^{ab}_{P;\mu \nu}$ and $\chi ^{ab}_P$ for two-gluons and ghost-antighost respectively as
\begin{equation}
\chi ^{ab}_{P;\mu \nu}(x,y)=\left< 0|\mathscr{A}^a_\mu (x) \mathscr{A}^b_\nu (y)|P_c\right>  , \quad
\chi ^{ab}_P (x,y)=\left< 0|\mathscr{C}^a(x)\bar{\mathscr{C}}^b(y)|P_c\right> .
\end{equation}
For the Yang-Mills theory, we define the amputated gluon BS amplitude $A$
and the amputated ghost BS amplitude $B$ by 
\begin{subequations}
\begin{align}
A^{\mu \nu}_{ab}(p;P)&:=\int \frac{d^Dq}{(2\pi )^D} \left( D^{(0)-1}\right) _{ab;a^\prime b^\prime }^{\mu \nu ;\mu^\prime \nu^\prime }(p,q;P)\bar{\chi}^{a^\prime b^\prime }_{P;\mu^\prime \nu^\prime }(q;P) , 
\\
B_{ab}(p;P)&:=\int \frac{d^Dq}{(2\pi )^D}\left( \Delta ^{(0)-1}\right) _{ab;a^\prime b^\prime }(p,q;P)\bar{\chi}^{a^\prime b^\prime }_P(q;P) .
\end{align}
\end{subequations}
\begin{figure}[t]
\centering
\includegraphics[scale=0.4]{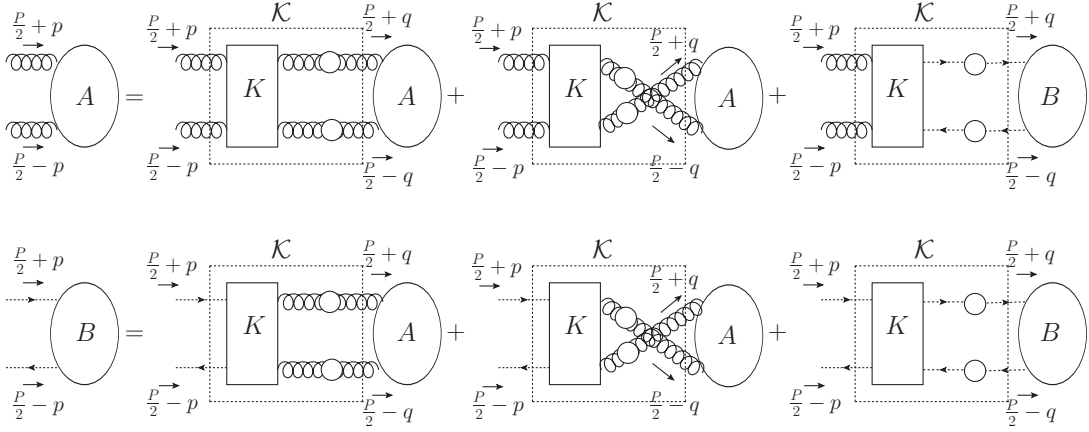}
\caption{
The homogeneous BS equation for the amputated gluon and ghost BS amplitudes for the Yang-Mills theory.
}
\label{fig:BS}
\end{figure}
%


\begin{figure}[t]
\centering
\includegraphics[scale=0.30]{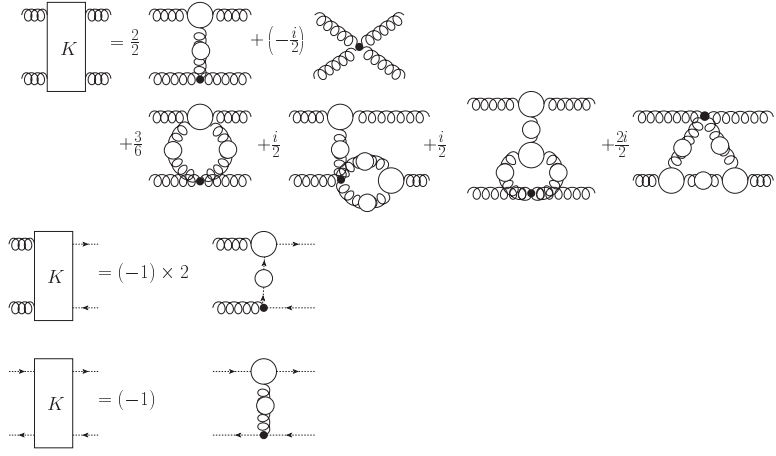}
\caption{
The kernels of the BS equation.
}
\label{fig:mYM-BS-kernel}
\end{figure}

Thus, the amputated BS amplitude obeys the homogeneous BS equation:
\begin{equation}
\begin{pmatrix}
A_{ab}^{\mu \nu}(p^+,p^-) \\ B_{ab}(p^+,p^-)
\end{pmatrix}
= \int \frac{d^Dq}{(2\pi )^D}
\begin{pmatrix}
\mathcal{K}^{ab;cd}_{\mu \nu ;\mu^\prime \nu^\prime } & \mathcal{K}^{ab;cd}_{\mu \nu ;} \\
\mathcal{K}^{ab;cd}_{\ \ ;\mu^\prime \nu^\prime } & \mathcal{K}^{ab;cd}
\end{pmatrix}
\begin{pmatrix}
A_{cd}^{\mu^\prime \nu^\prime }(q^+,q^-) \\ B_{cd}(q^+,q^-)  
\end{pmatrix} ,
\end{equation}
where we have defined the momenta $p^\pm$ by
\begin{align}
 p^\pm  :=p\pm \frac{P}{2} , \quad
 q^\pm  :=q\pm \frac{P}{2}  , 
\end{align}
and $\mathcal{K}$ by
\begin{align}
 \mathcal{K}^{ab;cd}_{\mu \nu ;\mu^\prime \nu ^\prime }=K^{ab;ef}_{\mu \nu ;\rho \sigma}D_{4}^{(0)}{}^{ef;cd}_{\rho \sigma ;\mu^\prime \nu^\prime } .
\end{align}
The diagram  of the homogeneous BS equation is given in Fig.~\ref{fig:BS}. 
Here the kernels of the BS equation are given as    Fig.~\ref{fig:mYM-BS-kernel}.
The explicit form of the kernel is given in Appendix A.
This BS equation can be compared with that obtained by 
 Meyers and Swanson \cite{MS13}.

\subsection{BS amplitude}

In what follows, we consider the color-singlet and scalar (i.e., spin zero) bound state.

Fukuda considered the decomposition for the amplitude:
\begin{align}
 A _{\mu \nu}(k;P)
=&  \left( g_{\mu \nu}-\frac{k _\mu k _\nu}{k^2} \right) A_1 +   \frac{k _\mu k _\nu}{k^2} A_2 + P_\mu P_\nu A_3 + (k_\mu P_\nu + k_\nu P_\mu) (k \cdot P) A_4 
\nonumber\\& 
+ (k_\mu P_\nu - k_\nu P_\mu) A_5 ,
\end{align}
where $A_1, ..., A_5$ are functions of $k^2, P^2$ and $(k \cdot P)^2$. 
Here $A_1$ and $A_2$ are respectively the transverse and longitudinal parts in terms of $k$, since we can define the transverse projector $t_{\mu \nu}$ and longitudinal projector $\ell_{\mu \nu}$ by
\begin{align}
t_{\mu \nu} &:= g_{\mu \nu}-\frac{k_\mu k_\nu}{k^2}  , \ 
\ell_{\mu \nu} :=\frac{k_\mu k_\nu}{k^2} , \
 \label{projector}
\end{align}
with the properties:
\begin{subequations}
\begin{align}
t_{\mu \nu} + \ell_{\mu \nu} =& g_{\mu \nu} , 
\\
t_{\mu \rho} t_{\rho \nu} =& t_{\mu \nu} , \
\ell_{\mu \rho} \ell_{\rho \nu} = \ell_{\mu \nu} , 
\\
t_{\mu \nu} =& t_{\nu\mu } , \ 
\ell_{\mu \nu} = \ell_{\nu\mu } , 
\\
 k_\mu t_{\mu \nu}=& 0=t_{\mu \nu}k_\nu ,
 \label{transverse}
 \\
 k_\mu \ell_{\mu \nu}=& k_\nu, \ \ell_{\mu \nu}k_\nu = k_\mu .
 \label{longitudinal}
\end{align}
\end{subequations}

In this paper, however, we regard the amplitude $A _{\mu \nu}$ as the function $A _{\mu \nu}(k^+;k^-)$ of $k^+:=k+P/2$ and $k^-:=k-P/2$, rather than the function $A _{\mu \nu}(k;P)$ of $k$ and $P$. 
For this purpose, this decomposition in terms of $k$ and $P$ is not suited in the presence of the non-vanishing total momentum $P^\mu$. 
We construct another decomposition of the amplitude in terms of $k^+$ and $k^-$. 
In particular, we require that the gluon BS amplitude $A_{\mu \nu}(k^+,k^-)$ is transverse  in the sense that 
\begin{equation}
 k^+_\mu A_{\mu \nu}=0=A_{\mu \nu}k^{-}_\nu .
\end{equation}
In view of these, we can define the modified transverse  projector in the presence of $P$ by
\begin{align}
T_{\mu \nu}&:= g_{\mu \nu}-\frac{k^-_\mu k^+_\nu}{k^+ \cdot k^-}   \ne 
T_{\nu\mu } := g_{\mu \nu}-\frac{k^+_\mu k^-_\nu}{k^+ \cdot k^-} 
 .
\end{align}
which is subject to
\begin{align}
k^+_\mu T_{\mu \nu} &=0=T_{\mu \nu} k^-_\nu .
\end{align}
Indeed, the modified transverse projector $T_{\mu \nu}$ reduces to the usual transverse projector (\ref{projector}) with the property (\ref{transverse}) in the limit $P \to 0$.

In the presence of $P$, we can construct another type of transverse projector. 
For this purpose, we can introduce the vectors $k^{+\bot }$ and $k^{-\bot  }$ which are orthogonal to $k^+$ and $k^-$ respectively:
\begin{align}
k^+\cdot k^{+\bot} &=0=k^-\cdot k^{-\bot}  .
\end{align}
Indeed, such vectors $k^{+\bot }$ and $k^{-\bot  }$ are constructed from $k^+$ and $k^-$ as 
\footnote{
This is obtained as follows.
We define the unit vectors $e^+$ and $e^-$ of $k^+$ and $k^-$: 
\begin{align}
e^+ :=\frac{k^+}{|k^+|} , \ 
e^- :=\frac{k^-}{|k^-|} ,
\end{align}
and we can use the ortho-normalization method due to Gram--Schmidt to construct the vectors $e^{+\bot}$ and $e^{-\bot}$ which are respectively orthogonal to $e^+$ and $e^-$:
\begin{align}
e^{\pm \bot}&=e^\mp - (e^\mp, e^\pm )e^\pm  
= \frac{k^\mp}{|k^\mp|}-\frac{k^\mp \cdot k^\pm }{|k^\mp||k^\pm |}\frac{k^\pm }{|k^\pm |}
= - \frac{k^\mp \cdot k^\pm }{|k^\mp||k^\pm |^2} \left( k^\pm  - \frac{ |k^\pm |^2}{k^\mp \cdot k^\pm } k^\mp \right) .
\end{align}
}
\begin{align}
k^{+\bot}_\mu :=k^+_\mu -\frac{|k^+|^2}{k^-\cdot k^+}k^-_\mu , \ 
k^{-\bot}_\mu :=k^-_\mu -\frac{|k^-|^2}{k^+\cdot k^-}k^+_\mu .
\end{align}

We can also define another transverse projector
\begin{align}
 L_{\mu \nu} :=\frac{k^{+\bot}_\mu k^{-\bot}_\nu}{k^{+\bot}\cdot k^{-\bot}} ,
\end{align}
with the property:
\begin{align}
k^+_\mu L_{\mu \nu}&=0=L_{\mu \nu}k^-_\nu .
\end{align}
Notice that $k^{+\bot }_\mu$ and $k^{-\bot  }_\mu$ vanish in the limit $P^\mu \to 0$. 
 In the limit $P \to 0$, therefore, $L_{\mu \nu}$ becomes ill-defined and does not reduce to $\ell_{\mu \nu}$ which is longitudinal to $k$. 
They satisfy the following properties:
\begin{subequations}
\begin{align}
T_{\mu \rho}T_{\rho \nu}&=g_{\mu  \nu}-\frac{k^-_{\mu } k^+_\nu}{k^-\cdot k^+} = T_{\mu \nu},
\\
L_{\mu \rho}L_{\rho \nu}&=-\frac{\left( k^+_{\mu } (k^- \cdot k^+) -{k^+}^2 k^-_{\mu}\right) \left({k^-}^2 k^+_{\nu }-k^-_{\nu } ({k^-}\cdot {k^+})\right)}{({k^-}\cdot {k^+})^3-{k^-}^2 {k^+}^2({k^-}\cdot {k^+})} ,
\\
T_{\mu \rho}L_{\rho \nu}&=-\frac{\left( k^+_{\mu } (k^-\cdot k^+)-{k^+}^2 k^-_{\mu}\right) \left( {k^-}^2 k^+_{\nu }-k^-_{\nu } (k^-\cdot k^+\right)}{(k^-\cdot k^+)^3-{k^-}^2 {k^+}^2 (k^-\cdot k^+)} ,
\end{align}
\end{subequations}
which lead to  
\begin{align}
T_{\mu \nu}T_{\nu \mu} =D-1 , \
L_{\mu \nu}L_{\nu \mu} =1 , \
T_{\mu \nu}L_{\nu \mu} =1 .
\label{TL-prop}
\end{align}

Now we proceed to construct the transverse amplitude. 
First, $k^+_\mu A_{\mu \nu}=0$ is satisfied by the tensor of the form:
\begin{equation}
A_{\mu \nu}=f  T_{\mu \nu} +g _\nu  k^{+\bot}_\mu .
\end{equation}
Second, the condition  $A_{\mu \nu}k^{-}_\nu =0$ yields 
\begin{equation}
A_{\mu \nu}k^-_\nu 
=f   T_{\mu \nu}k^-_\nu +g _\nu k^{+\bot}_\mu k^-_\nu
=( g _\nu   k^-_\nu ) k^{+\bot}_\mu =0 .
\end{equation}
 $g _\nu  k^-_\nu =0$ follows from $g _\nu   =g k^{-\bot}_\nu$.
Thus, $A_{\mu \nu}$ has the form:
\begin{align}
A_{\mu \nu}&
=f T_{\mu \nu} +h  L_{\mu \nu}.
\end{align}
Consequently, we define the gluon BS amplitude $A_{\mu \nu}$ by \begin{equation}
A_{\mu \nu}(k^+,k^-)=A_1(k^+,k^-)T_{\mu \nu}+A_2(k^+,k^-)L_{\mu \nu} .
\label{projection}
\end{equation}
Multiplying both sides of $A_{\mu \nu}(k^+,k^-)$ with 
$g_{\nu \mu}$ and $L_{\nu \mu}$ from the left, we obtain using (\ref{TL-prop}):
\begin{align}
g_{\nu \mu} A_{\mu \nu}(k^+,k^-)&=(D-1)A_1(k^+,k^-)+A_2(k^+,k^-) ,
\nonumber\\
L_{\nu \mu} A_{\mu \nu}(k^+,k^-)&=A_1(k^+,k^-)+A_2(k^+,k^-) ,
\end{align}
which has the matrix form:
\begin{equation}
\begin{pmatrix}
g_{\nu \mu} \\ L_{\nu \mu}
\end{pmatrix}
A_{\mu \nu}(k^+,k^-)=
\begin{pmatrix}
D-1 & 1 \\
1 & 1
\end{pmatrix}
\begin{pmatrix}
A_1(k^+,k^-) \\ A_2(k^+,k^-)
\end{pmatrix} .
\label{IM}
\end{equation}
This is solved with respect to $A_1,A_2$:
\begin{equation}
\begin{pmatrix}
A_1(k^+,k^-) \\ A_2(k^+,k^-)
\end{pmatrix}
=
\begin{pmatrix}
\frac{1}{D-2} & \frac{1}{2-D} \\
\frac{1}{2-D} & \frac{D-1}{D-2} \\
\end{pmatrix}
\begin{pmatrix}
g_{\nu \mu} \\ L_{\nu \mu}
\end{pmatrix}
A_{\mu \nu}(k^+,k^-) .
\end{equation}
Thus, $A_1$ and $A_2$ are extracted from $A_{\mu \nu}$ by operating the projection operators $P^{A_1}_{\nu \mu}$ and $P^{A_2}_{\nu \mu}$ respectively:
\begin{subequations}
\begin{align}
 A_1 = P^{A_1}_{\nu \mu} A_{\mu \nu}, \ P^{A_1}_{\nu \mu}&:=\frac{1}{D-2}g_{\nu \mu}+\frac{1}{2-D}L_{\nu \mu} ,
\\
A_2 = P^{A_2}_{\nu \mu} A_{\mu \nu}, \ P^{A_2}_{\nu \mu}&:=\frac{1}{2-D}g_{\nu \mu}+\frac{D-1}{D-2} L_{\nu \mu} .
\end{align}
\end{subequations}

Consequently, the homogeneous BS equation for the gluon amplitude is rewritten 
using the abbreviated form:
\begin{subequations}
\begin{align}
A^{ab}_1(k^+,k^-)&=i\int \frac{d^Dq}{(2\pi )^D}P^{A_1}_{\nu \mu}\left( \mathcal{K}^{ab;cd}_{\mu \nu ;\mu '\nu '}A_{\mu '\nu '}^{cd}(q^+,q^-)+\mathcal{K}^{ab;cd}_{\mu \nu ;}B^{cd}(q^+,q^-)\right) ,
\\
A^{ab}_2(k^+,k^-)&=i\int \frac{d^Dq}{(2\pi )^D}P^{A_2}_{\nu \mu}\left( \mathcal{K}^{ab;cd}_{\mu \nu ;\mu '\nu '}A_{\mu '\nu '}^{cd}(q^+,q^-)+\mathcal{K}^{ab;cd}_{\mu \nu ;}B^{cd}(q^+,q^-)\right) .
\end{align}
\end{subequations}
The homogeneous BS equation for the gluon and ghost amplitudes is obtained
\begin{equation}
\begin{pmatrix}
A^{ab}_1(p^+,p^-) \\ A^{ab}_2(p^+,p^-) \\ B^{ab}(p^+,p^-)
\end{pmatrix}
=i\int \frac{d^Dq}{(2\pi )^D}
\begin{pmatrix}
a_{11} & a_{12} & a_{13} \\
a_{21} & a_{22} & a_{23} \\
a_{31} & a_{32} & a_{33} \\
\end{pmatrix}^{ab;cd}
\begin{pmatrix}
A_1^{cd}(q^+,q^-) \\ A_2^{cd}(q^+,q^-) \\ B^{cd}(q^+,q^-)
\end{pmatrix} ,
\label{BSE}
\end{equation}
where the matrix element is given as 
\begin{subequations}
\begin{align}
P^{A_1}_{\nu \mu}\left( \mathcal{K}^{ab;cd}_{\mu \nu ;\mu '\nu '}A^{cd}_{\mu '\nu '}+\mathcal{K}^{ab;cd}_{\mu \nu ;}B^{cd}\right) &=a_{11}A_1+a_{12}A_2+a_{13}B ,
\\
P^{A_2}_{\nu \mu}\left( \mathcal{K}^{ab;cd}_{\mu \nu ;\mu '\nu '}A^{cd}_{\mu '\nu '}+\mathcal{K}^{ab;cd}_{\mu \nu ;}B^{cd}\right) &=a_{21}A_1+a_{22}A_2+a_{23}B ,
\\
\left( \mathcal{K}^{ab;cd}_{\ \ ;\mu '\nu '}A^{cd}_{\mu '\nu '}+\mathcal{K}^{ab;cd}B^{cd}\right) &=a_{31}A_1+a_{32}A_2+a_{33}B .
\end{align}
\end{subequations}
For the explicit forms of the kernels, see Appendix~\ref{section:kernel-general}. 



\section{The coupled BS equation in the massless case}

In this section, we study the massless case $m=0$ to reproduce the Fukuda's results by correcting some errors \cite{Fukuda78} in our framework, before studying the massive case $m \ne 0$ in the subsequent sections.
We obtain the numerical solutions of the amputated gluon and ghost BS amplitudes for the coupled homogeneous BS equations in the massless case $m=0$. 
To write down the manageable integral equations, 
we perform the Wick rotation to the relative momenta $p$ and $q$ to obtain the Euclidean momenta  $p_{E}$ and $q_{E}$ \cite{Wick54}. 
For the total momentum $P:= (E, \bm{P})$, however, we restrict our analysis to the vanishing total momentum $P^\mu =0$. 
Although this is unrealistic for the bound state,  
this investigation is a preliminary step for studying the true bound sates dictated by $P^\mu \ne 0$ in the subsequent sections. 
Even  the special result obtained at $P^\mu =0$ is expected to give some helpful information on the mass spectrum $P^2$, provided that the Yang-Mills theory has some similarity to the Wick-Cutkosky model \cite{Cutkosky54,Silagadze98}. 
In what follows, we adopt the Landau gauge in the Lorenz gauge fixing. 


We replace the full propagators and the full vertices by the bare propagators and bare vertices in the BS equation. 
Then the homogeneous BS equation reduces to the integral equation of the Fredholm type \cite{J.Kondo59}.

First of all, we assume that the gauge coupling constant $g$ does not run, i.e., is standing.
After the angular integration in the integration measure,   
 the homogeneous BS equation for the $SU(N)$ Yang-Mills theory in the Landau gauge at $P^\mu =0$ is cast into the coupled integral equation of the form:
\begin{equation}
 \begin{pmatrix}
A_1(p_{E}^2) \\ B(p_{E}^2)
 \end{pmatrix}
=C \int_{\Lambda _I^2}^{\Lambda _U^2} \frac{dq_{E}^2}{q_{E}^2}
 \begin{pmatrix}
a_{11}(p_{E}^2,q_{E}^2) & a_{12}(p_{E}^2,q_{E}^2) \\
a_{21}(p_{E}^2,q_{E}^2) & a_{22}(p_{E}^2,q_{E}^2) \\
 \end{pmatrix}
 \begin{pmatrix}
A_1(q_{E}^2) \\ B(q_{E}^2)
 \end{pmatrix} ,
\quad  C=\frac{Ng^2}{16\pi ^2} , 
\label{BS-coupled1}
\end{equation}
with the kernel written in the matrix with the elements:
\begin{subequations}
\begin{align}
 {\text{Gluon--Gluon:}}~a_{11}&=\left\{ \frac{3}{2 } + \frac{22}{3}\frac{q_{E}^2}{p_{E}^2} \right\} \theta (p_{E}^2-q_{E}^2)+\left\{  {2} + 8\frac{ p_{E}^2}{q_{E}^2} -\frac{7}{6}\frac{p_{E}^4}{q_{E}^4} \right\} \theta (q_{E}^2-p_{E}^2) >0 ,
\\
 {\text{Gluon--Ghost:}}~a_{12}&=\left\{ - \frac{1}{3}\frac{q_{E}^2}{p_{E}^2}\right\} \theta (p_{E}^2-q_{E}^2)+\left\{ - \frac{ 2}{3} +  \frac{1}{3}\frac{p_{E}^2}{ q_{E}^2}\right\} \theta (q_{E}^2-p_{E}^2) <0 ,
\\
 {\text{Ghost--Gluon:}}~a_{21}&=\left\{   - \frac{3}{4} + \frac{1}{4} \frac{q_{E}^2}{p_{E}^2}   \right\} \theta (p_{E}^2-q_{E}^2)+ \left\{ 
   - \frac{3}{4} \frac{p_{E}^2}{q_{E}^2} + \frac{1}{4} \frac{p_{E}^4}{q_{E}^4}    \right\} \theta (q_{E}^2-p_{E}^2) <0 ,
\\
 {\text{Ghost--Ghost:}}~a_{22}&=\left\{ - \frac{3}{4} \frac{q_{E}^2}{ p_{E}^2}\right\} \theta (p_{E}^2-q_{E}^2)+\left\{ -\frac{3}{4} \frac{ p_{E}^2}{ q_{E}^2}\right\} \theta (q_{E}^2-p_{E}^2) <0 .
\end{align}
\label{BS-coupled1-kernel}
\end{subequations}
%
%
%
%
\begin{figure}[t]
\begin{minipage}{0.48 \hsize}
\centering
\includegraphics[scale=0.60]{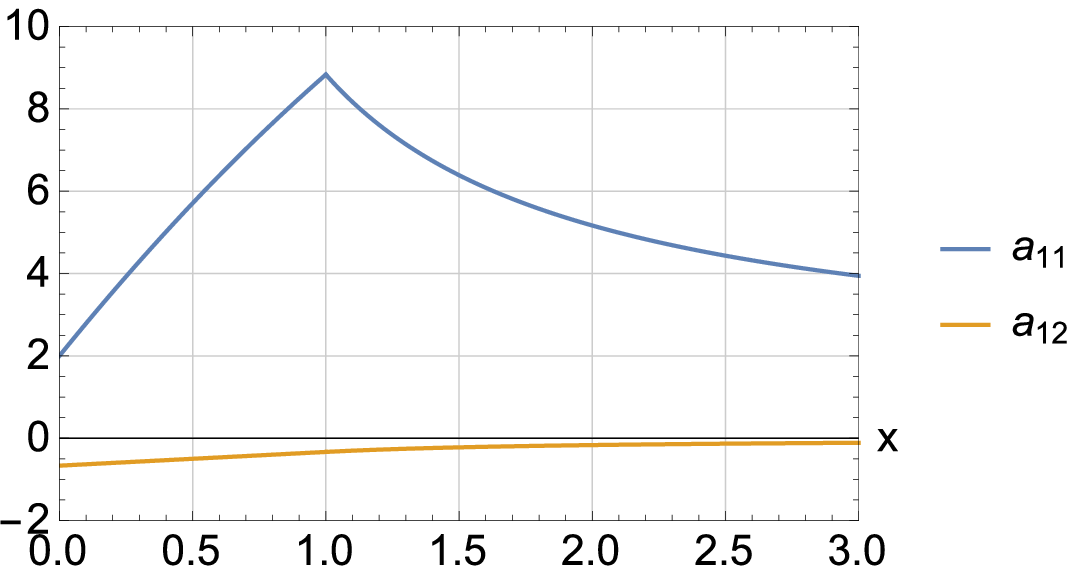}
\end{minipage}
\begin{minipage}{0.48 \hsize}
\centering
\includegraphics[scale=0.60]{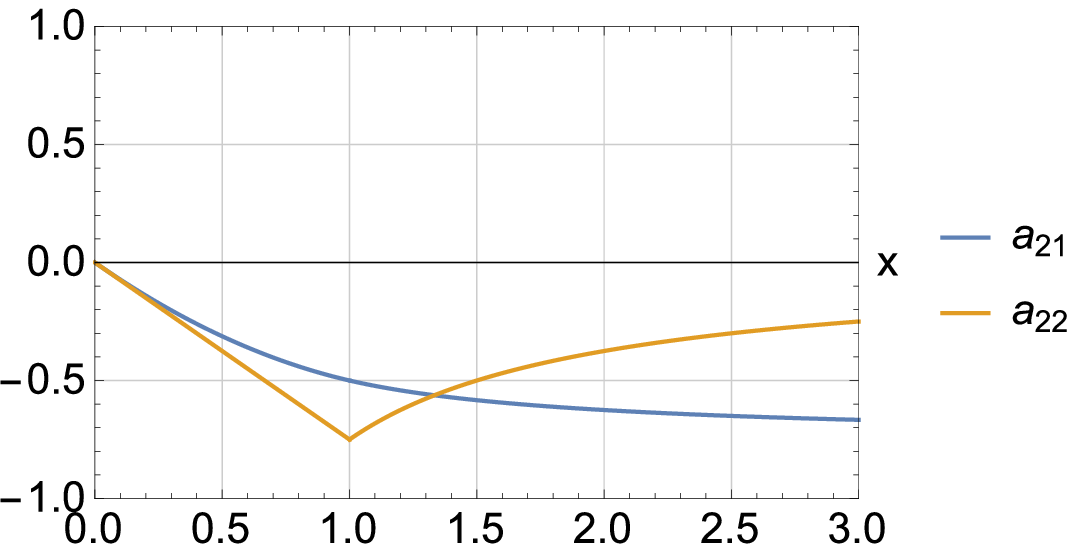}
\end{minipage}
\caption{
The matrix elements $a_{11},a_{12}$ and $a_{21},a_{22}$ of the kernel as functions of $x :=\frac{p^2}{q^2}$.
}
\label{fig:kernel-elements}
\end{figure}
Notice that the integration measure is $dq^2/q^2$, modified from the naive $dq^2$, which affects specifying the kernel.  It turns out that this choice is efficient for the bound state as shown in the next section. 
Fig.~\ref{fig:kernel-elements} is the plot of the matrix elements  $a_{jk}$ of the kernel as  functions of $x :=\frac{p_{E}^2}{q_{E}^2}$.
Only $a_{11}$ is positive, while the other elements are all negative. For the magnitude, the gluon-gluon component $a_{11}$ is large compared to the other elements. 
This fact suggest that it is enough to include only the contribution of $a_{11}$ to obtain the qualitatively correct result for the bound state. 
Therefore, it is expected that the positively large element $a_{11}$ dominantly produces the attractive force between two gluons to form the 2-body bound state. 
This observation agrees with the result of Fukuda \cite{Fukuda78}, but the kernel elements are slightly different from those given in \cite{Fukuda78}. 
See Appendix~\ref{section:kernel-Landau_P=0} for more details. 

\begin{figure}[h]
\includegraphics[scale=0.55]{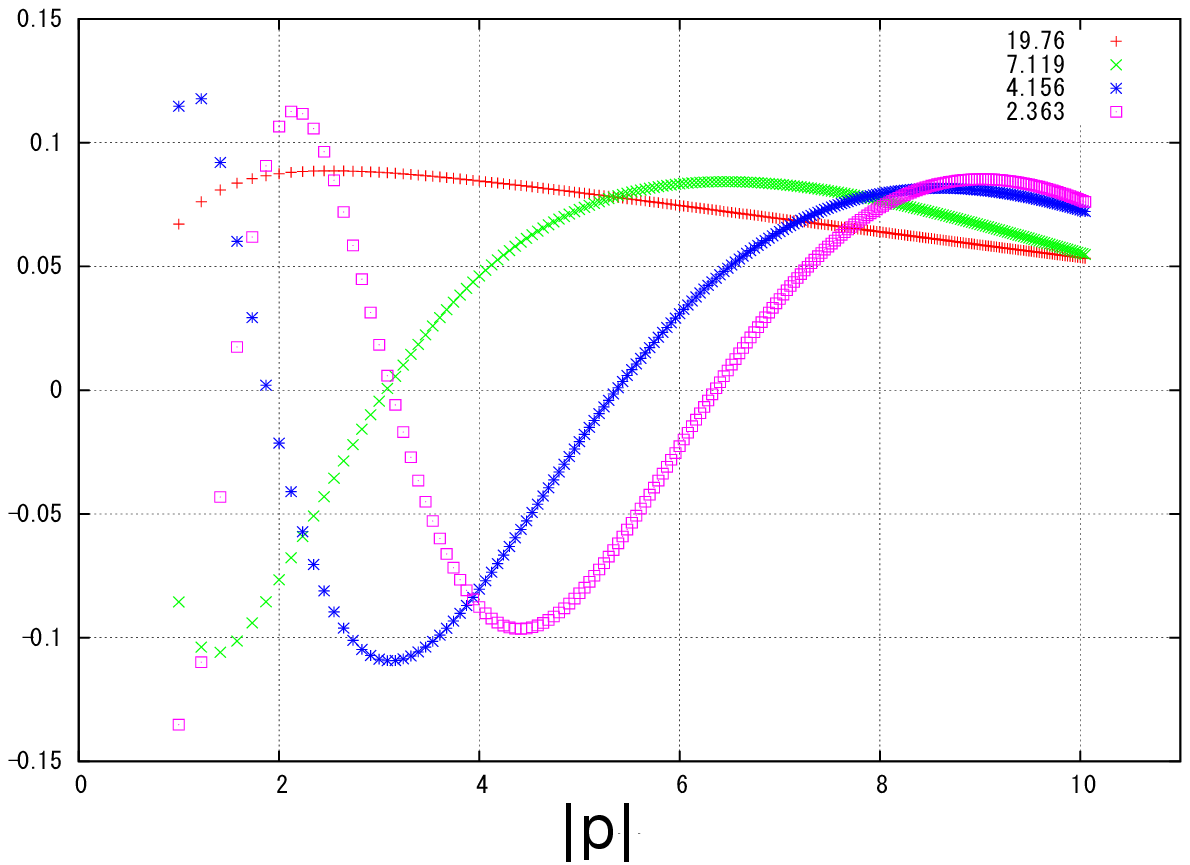}
\quad\quad
\includegraphics[scale=0.55]{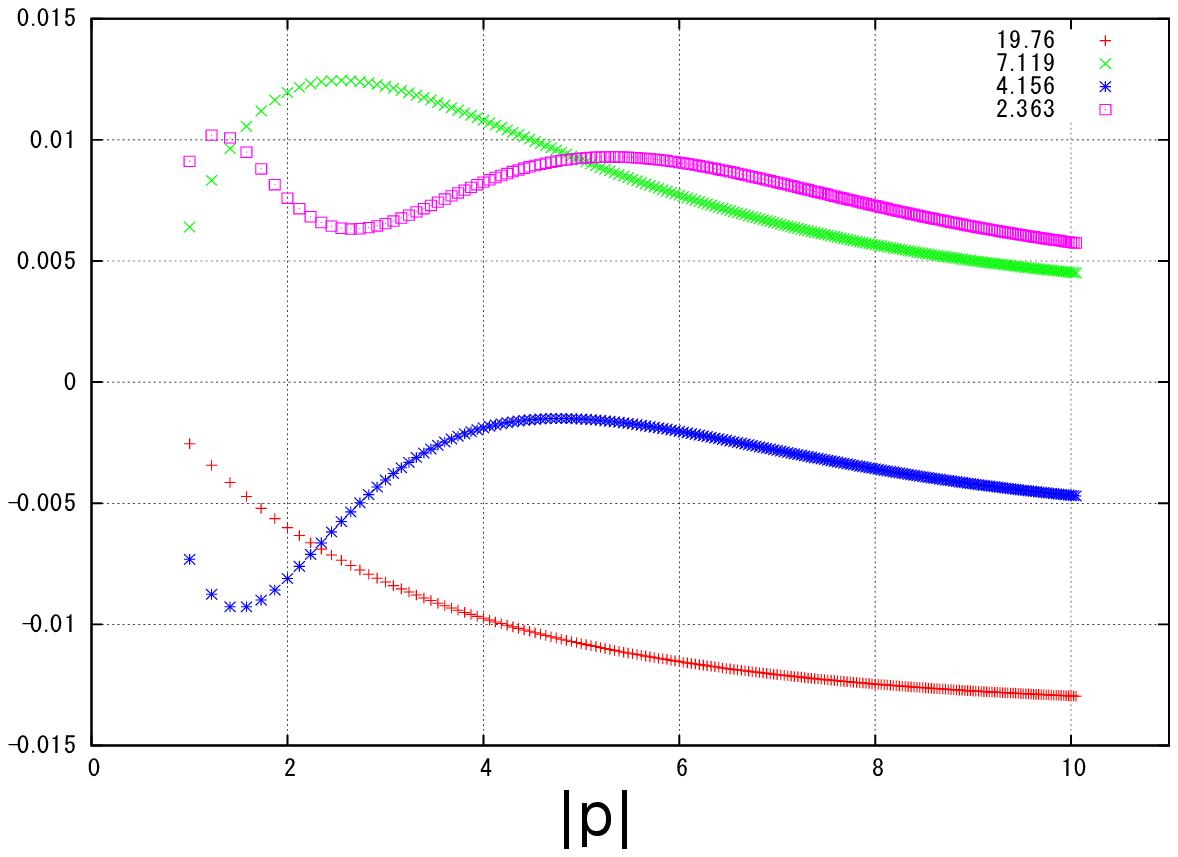}
\caption{
(Left) Gluon amplitude $A_1$,
(Right) Ghost amplitude $B$  
obtained by solving the integral equation when $a_{11}\not =0, a_{21}\not =0, a_{12}=a_{22}=0$.
Here we adopt $\Lambda_I^2=1$, $\Lambda_U^2=101$ and 200 partitions for obtaining numerical solutions.
The red lines correspond  to the largest eigenvalue of $1/C$.
}
\label{fig:A1-B-1}
\end{figure}

First, we take into account only the element $a_{11}$ by putting $a_{12} =0$ 
to solve the integral equation for the gluon amplitude $A_1$ alone.
The result for the gluon amplitude $A_1$ is given in the left panel of Fig.~\ref{fig:A1-B-1}.
Then the ghost amplitude $B$ is obtained by taking into account $a_{21}$ by putting $a_{22} =0$ from the solution for $A_1$. 
The result for the ghost amplitude $B$ obtained in this way is given in the right panel of Fig.~\ref{fig:A1-B-1}.

The BS equation is regarded as the eigenvalue equation and a BS amplitude as a solution of the BS equation is identified with an eigenfunction  associated to each eigenvalue of $1/C$.  
To perform the numerical calculations, we need to divide the integration region $[\Lambda_I^2,\Lambda_U^2]$ into a sufficient number of small steps. Then the integral equation has the same number of eigenvalues (and the associated eigenfunctions) as the number of partitions.
Here we have adopted the infrared and ultraviolet cutoff to be $\Lambda_I^2=1$ and $\Lambda_U^2=101$ respectively, and tried to divide the interval $[\Lambda_I^2,\Lambda_U^2]$ to 10, 100, and 200 pieces to see whether the result  converges or not as the number of partition increases. 
The results are shown in Table~\ref{Table:eigen-1}.
 Fig.~\ref{fig:A1-B-1} is the plot of the eigenfunction  corresponding to an eigenvalue. 
Here we present only the graphs obtained from the result of 200 partitions, which  gives the convergent result. 
For this choice of partitions, indeed, we observe that the largest eigenvalue of $1/C$ and the associated eigenfunction  is convergent as confirmed according to Table~\ref{Table:eigen-1}, which are denoted by the red curves in Fig.~\ref{fig:A1-B-1}.
The largest eigenvalue of $1/C$ corresponds to the nodeless eigenfunction, and the node of the eigenfunction monotonically increases as the eigenvalue decreases. 
Here we define the coupling constant for $SU(2)$ and $SU(3)$ as follows. 
\begin{equation}
 C=\frac{N g^2}{16\pi ^2}=\frac{N}{4\pi}\times \frac{g^2}{4\pi}=\frac{N}{4\pi} \alpha , 
\quad \alpha := \frac{g^2}{4\pi} \to \alpha_{\text{SU(2)}}=\frac{4\pi C}{2}=\frac{2\pi}{\frac{1}{C}} ,
\quad \alpha_{\text{SU(3)}}=\frac{4\pi C}{3}=\frac{4\pi}{3\frac{1}{C}} .
\end{equation}
 
\begin{table}[h]
\centering
\caption{
A set of eigenvalues for $\frac{1}{C}$ in the descending order obtained for various  partitions.
}
\begin{tabular}{|c|c|c|c|c|c|c|c|} \hline
 number of partitions & \multicolumn{7}{c|}{$\frac{1}{C}$} \\ \hline
10 & 46.14 & 13.17 & 4.756 & 2.063 & 0.9797 & 0.5191 & ...   \\ \hline
100 & 19.86 & 7.180 & 4.301 & 2.575 & 1.802 & 1.249 & ...  \\ \hline
200 & 19.76 & 7.119 & 4.156 & 2.363 & 1.544 & 1.050 & ...  \\ \hline
\end{tabular}
\label{Table:eigen-1}
\end{table}
 
\begin{figure}[t]
\includegraphics[scale=0.55]{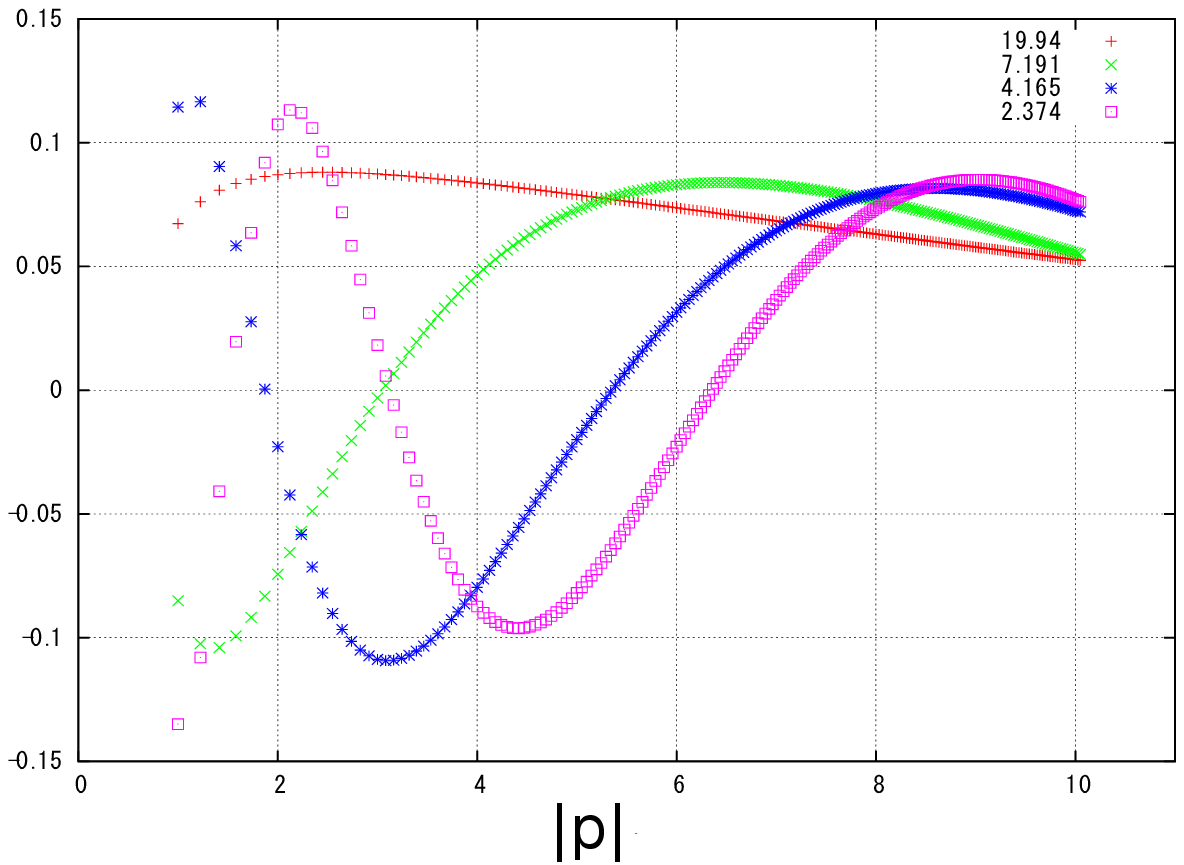}
\quad\quad
\includegraphics[scale=0.55]{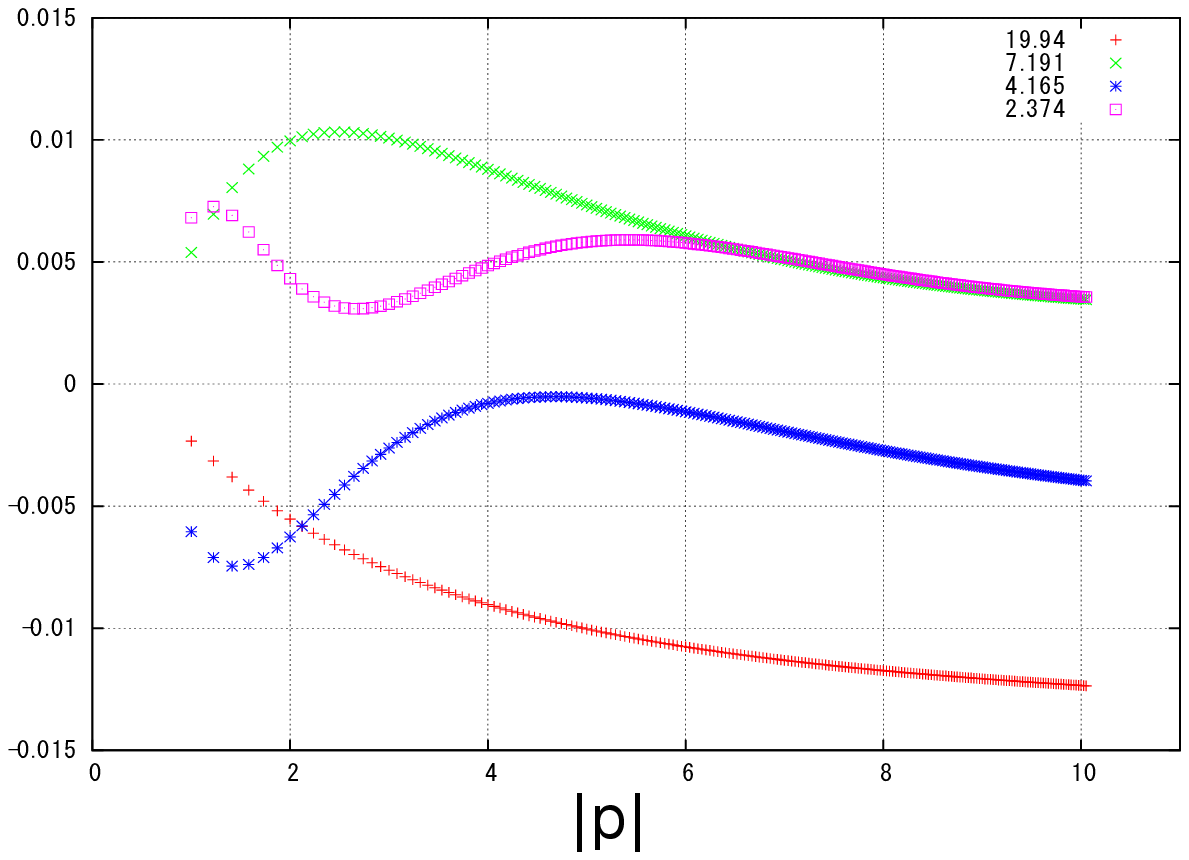}
\caption{
(Left) Gluon amplitude $A_1$,
(Right) Ghost amplitude $B$  
obtained by solving the integral equation when all $a_{jk}\not =0$.
Here we adopt $\Lambda_I^2=1$, $\Lambda_U^2=101$ and 200 partitions for obtaining numerical solutions.
The red lines correspond  to the largest eigenvalue of $1/C$.
}
\label{fig:A1-B-2}
\end{figure}

Second, we take into account all the matrix elements. 
The numerical solutions are given in Fig.~\ref{fig:A1-B-2}. 
See also Table~\ref{Table:eigen-2}.
Fig.~\ref{fig:A1-B-2} shows no sizable difference from Fig.~\ref{fig:A1-B-1}. This confirms the above observation is correct also from the quantitative point of view: 
It is enough to include the gluon-gluon contribution $a_{11}$ to study the gluon BS amplitude $A_1$.

\begin{table}[t]
\centering
\caption{
A set of eigenvalues for $\frac{1}{C}$ in the descending order obtained for various  partitions.
}
\begin{tabular}{|c|c|c|c|c|c|c|c|} \hline
 number of partitions & \multicolumn{7}{c|}{$\frac{1}{C}$} \\ \hline
10 & 46.38 & 13.20 & 4.770 & $-$3.864 & 2.063 & $-$1.234 & ...  \\ \hline
100 & 20.04 & 7.255 & 4.311 & 2.588 & 1.803 & $-$1.456 & ...  \\ \hline
200 & 19.94 & 7.191 & 4.165 & 2.374 & 1.546 & $-$1.450 & ...  \\ \hline
\end{tabular}
\label{Table:eigen-2}
\end{table}

The solution can also be analyzed by converting the integral equation to the equivalent differential equation, which is simplified  by taking into account the dominant element $a_{11}$ alone.
The resulting equation becomes the fourth-order differential equation with the boundary conditions imposed at the infrared lower bound and the ultraviolet upper bound.  
In fact, the solution is analyzed in \cite{Fukamachi15}, but the details are omitted here.

\section{Gluon-gluon BS equation in the massive case}

In what follows, we take into account the gluon-gluon contribution alone in the BS equation.
Then the  homogeneous BS equation for the  {amputated BS amplitude} $D_{4}^{(0)}{}^{-1}\bar\chi$ reads
\begin{equation}
\Gamma ^{(0) ac}_{\mu \rho}(p^+)\Gamma ^{(0) bd}_{\nu \sigma}(p^-) \bar{\chi}_{cd}^{\rho \sigma}(p;P)
= -\int \frac{d^4q}{(2\pi )^4} \left[ 2K_3\,^{ab;cd}_{\mu \nu ;\rho \sigma}(p,q;P)-i K_4\,^{ab;cd}_{\mu \nu ;\rho \sigma}(p,q;P) \right] \bar{\chi} _{cd}^{\rho \sigma}(q;P) ,
\label{BSE2}
\end{equation}
where 
%
%
 $\Gamma ^{(0)ab}_{\mu \nu}$ is the inverse propagator for the massive gluon defined by
\begin{align}
\Gamma ^{(0)ab}_{\mu \nu}(p)&= \delta ^{ab} \left[ -\left( p^2-m^2\right) g_{\mu \nu}+\left( 1-\frac{1}{\alpha}\right) p_\mu p_\nu \right] ,
%
%
%
%
\end{align}
$K_3$ and $K_4$ are the kernels defined by
\begin{align}
K_3\,^{ab;cd}_{\mu \nu ;\rho \sigma}(p,q;P) &:=\Gamma ^{(0)ace}_{\mu \rho \alpha}(-p^+,q^+,k^-)D^{(0)ee^\prime}_{\alpha \alpha^\prime}(k^-) \Gamma ^{(0)bde^\prime}_{\nu \sigma \alpha ^\prime}(p^-,-q^-,-k^-) ,
\nonumber\\
K_4\,^{ab;cd}_{\mu \nu ;\rho \sigma}(p,q;P) &:=\Gamma ^{(0)abdc}_{\mu \nu \sigma \rho}(-p^+,p^-,-q^-,q^+) ,
\quad
k^- :=p-q .
\end{align}
Here  $\Gamma ^{(0)ace}_{\mu \rho \alpha}$ is the three-gluon vertex and  $\Gamma ^{(0)abdc}_{\mu \nu \sigma \rho}$ is the four-gluon vertex,  
while $D^{(0)ab}_{\mu \nu}$ is the ordinary massless gluon propagator defined by
\begin{align}
D^{(0)ab}_{\mu \nu}(p)&= \delta ^{ab} \frac{-i }{p^2}\left[ g_{\mu \nu}-(1-\alpha )\frac{p_\mu p_\nu}{p^2}  \right] .
\end{align}
It should be remarked that we adopt the massless gluon (with the propagator $D^{(0)ab}_{\mu \nu}(p)$) to mediate the strong interactions, while the constituent gluons are massive with the mass $m$. 
This approximation will be allowed to study the large and intermediate momentum region except the very small momentum region where the gluon propagator strongly depends on the approximation adopted. 
It is possible to replace the massless gluon mediating the strong interactions by the massive gluon with the same mass $m$ as the constituent gluon, which however leads to the very complicated expression for the BS kernel. 
This issue will be tackled in a subsequent paper \cite{FKNS}.

In what follows, we restrict our consideration to the BS amplitude $\bar{\chi}^{ab}_{\mu \nu}(p;P)$ of the transverse type:
\begin{equation}
 \bar{\chi}^{ab}_{\mu \nu}(p;P)= \delta ^{ab} \left( g_{\mu \nu}-\frac{p^-_\mu p^+_\nu}{p^+\cdot p^-}\right) f(p;P) .
\end{equation}
Indeed, this amplitude is transverse in the sense that
\begin{equation}
p_+^\mu \bar{\chi}^{ab}_{\mu \nu}(p;P)= 0 
= \bar{\chi}^{ab}_{\mu \nu}(p;P) p_-^\nu .
\end{equation}
Therefore, the left-hand side of (\ref{BSE2}) does not depend on the parameter $\alpha$ and reads 
\begin{equation}
 \Gamma ^{(0) ac}_{\mu \rho}(p^+)\Gamma ^{(0) bd}_{\nu \sigma}(p^-) \bar{\chi}_{cd}^{\rho \sigma}(p;P)
= \delta ^{ab} \frac{({p^-}^2-m^2)({p^+}^2-m^2)\left( p^-\cdot p^+ g_{\mu \nu}-p^-_\mu p^+_\nu \right)}{p^-\cdot p^+} f(p;P) .
\label{left}
\end{equation}
By taking the contraction on the Lorentz indices, the left-hand side reads 
\begin{align}
&
g^{\mu \nu} \Gamma ^{(0) ac}_{\mu \rho}(p^+)\Gamma ^{(0) bd}_{\nu \sigma}(p^-) \bar{\chi}_{cd}^{\rho \sigma}(p;P)
=3 \delta ^{ab} ({p^-}^2-m^2)({p^+}^2-m^2)f(p;P) .
\label{BS-left}
\end{align}
By adopting the Landau gauge $\alpha=0$, the integrand of the right-hand side of (\ref{BSE2}) reads
\begin{subequations}
\begin{align}
&g^{\mu \nu} K_3\,^{ab;cd}_{\mu \nu ;\rho \sigma}(p,q;P) \bar{\chi}_{cd}^{\rho \sigma}(q;P) 
\notag \\
&= - \delta ^{ab} \frac{i g^2 f(q;P)}{{k^-}^4 {q^-}\cdot {q^+}}
\Big\{ 5 {k^-}^4 {q^-}\cdot {q^+}-{k^-}^2 \Big[ {k^-}\cdot {q^+} \left({p^-}\cdot {q^-}+2{p^+}\cdot {q^-}+{q^-}^2+2 {q^-}\cdot {q^+}\right) 
\notag \\
&\hspace{3cm} +{k^-}\cdot {q^-} \left( 2 {k^-}\cdot {q^+}+2{p^-}\cdot {q^+}+{p^+}\cdot {q^+}+2 {q^-}\cdot{q^+}+{q^+}^2\right) 
\notag \\
&\hspace{3cm} -3 {k^-}\cdot {p^-} {q^-}\cdot{q^+}-3 {k^-}\cdot {p^+} {q^-}\cdot {q^+}+{p^-}\cdot{q^+} {p^+}\cdot {q^-}-4 {p^-}\cdot {p^+} {q^-}\cdot{q^+} \notag \\
&\hspace{3cm} -{q^+}^2 {p^-}\cdot {q^-}-2 {p^-}\cdot {q^+}{q^-}\cdot {q^+}-{q^-}^2 {p^+}\cdot {q^+}-2{p^+}\cdot {q^-} {q^-}\cdot {q^+}-2 {q^-}^2{q^+}^2
\notag \\
&\hspace{3cm} -3 ({q^-}\cdot {q^+})^2\Big] 
\notag \\
&\hspace{3cm} -{k^-}\cdot {p^+}\left[{k^-}\cdot {p^-} {q^-}\cdot {q^+}+{k^-}\cdot{q^-} ({p^-}\cdot {q^+}+2 {q^-}\cdot {q^+})+{q^-}^2{k^-}\cdot {q^+}\right] 
\notag \\
&\hspace{3cm} -{k^-}\cdot {p^-}\left[{k^-}\cdot {q^+} ({p^+}\cdot {q^-}+2 {q^-}\cdot{q^+})+{q^+}^2 {k^-}\cdot {q^-}\right]-{q^-}^2({k^-}\cdot {q^+})^2
\notag \\
&\hspace{3cm} -{q^+}^2 ({k^-}\cdot{q^-})^2\Big\} ,
\\
& -i g^{\mu \nu} K_4\,^{ab;cd}_{\mu \nu ;\rho \sigma}(p,q;P) \bar{\chi}_{cd}^{\rho \sigma}(q;P)
= 18 i g^2 f(q;P) \delta ^{ab} .
\end{align}
\label{BS-right}
\end{subequations}

\section{gluon-gluon BS amplitude}

We take the approximation in which the angular dependence between the total momentum $P$ and the relative momenta $p, q$, namely, the terms $P\cdot p$ and  $P\cdot q$ are neglected. Then the left-hand side of (\ref{BS-left}) reduces to
\begin{align}
 & g^{\mu \nu} \Gamma ^{(0) ac}_{\mu \rho}(p^+)\Gamma ^{(0) bd}_{\nu \sigma}(p^-) \bar{\chi}_{cd}^{\rho \sigma}(p;P)  
\nonumber\\
&=3 \delta ^{ab} ({p^-}^2-m^2)({p^+}^2-m^2)f(p;P) .
\nonumber\\
&=   \frac{3}{16}\left[ 8 p^2 \left( P^2-4m^2 \right) +\left( P^2-4m^2\right) ^2+16p^4-16 (p\cdot P)^2\right] f(p;P) \delta ^{ab}
\nonumber\\
&\rightarrow 3\left( p^2+\frac{P^2}{4}-m^2\right)^2 f(p ;P ) \delta ^{ab} .
\end{align}
In the same approximation as that for the left-hand side  of (\ref{BS-left}), the right-hand side of (\ref{BS-right}) reduces to
\begin{subequations}
\begin{align}
 & g^{\mu \nu}K_3\,^{ab;cd}_{\mu \nu ;\rho \sigma}(p,q;0) \bar{\chi}_{cd}^{\rho \sigma}(q;0) 
\nonumber\\
&=
\delta ^{ab} \frac{ig^2 f(q;P)}{8(q^2-\frac{P^2}{4})(p-q)^4} 
\Big\{ 28 p^4 \left(P^2-4 q^2\right)+\left(88 p^2+90 P^2-32 q^2\right) (p\cdot q)^2
\notag \\
& +p^2 \left(-7 P^4+72 P^2 q^2-280 q^4\right) 
+2 \left[ p^2 \left(152 q^2-52 P^2\right)+7 P^4-62 P^2 q^2+144 q^4 \right] p\cdot q 
\notag \\
& -160 (p\cdot q)^3-7 P^4 q^2+38 P^2
 q^4-96 q^6 \Big\} ,
\\
& -i g^{\mu \nu} K_4\,^{ab;cd}_{\mu \nu ;\rho \sigma}(p,q;P) \bar{\chi}_{cd}^{\rho \sigma}(q;P)
= 18 i g^2 f(q;P) \delta ^{ab} .
\end{align}
\end{subequations}
Here notice that 
$g^{\mu \nu} K_4\,^{ab;cd}_{\mu \nu ;\rho \sigma}(p,q;P) \bar\chi^{cd}_{\rho \sigma}(q;P)$ does not depend on $P$ and agrees with the original expression (\ref{BS-right}). 

We perform the Wick rotation  to the Euclidean region for performing the momentum integration explicitly.  
Notice that we apply the Wick rotation to $p$ and $q$ to convert them to the Euclidean momenta $p_{E}$ and $q_{E}$, while we leave the total momentum $P$ the Minkowski momentum. 
After performing the integration over the angular variables in the integration measure $d^4q_{E}$ for the Euclidean momentum $q_{E}$, we obtain the integral equation with respect to the magnitude $\sqrt{q_{E}\cdot q_{E}}$ of the Euclidean momentum:
\begin{subequations}
\begin{align}
 & 3\left( p_{E}^2-\frac{P^2}{4}+m^2 \right)^2  f(p_{E}^2;P^2) =  C\int q_{E}^2 dq_{E}^2 K(p_{E}^2,q_{E}^2;P) f(q_{E}^2;P^2) ,
\nonumber\\
 & K(p_{E}^2,q_{E}^2;P^2) =  \frac{1}{P^2+4q_{E}^2}\left( \frac{7 P^4}{p_E^2}+\frac{37 P^2 q_E^2}{2 p_E^2}+\frac{88 q_E^4}{p_E^2}+10 P^2+18 q_E^2\right) \theta (p_{E}^2-q_{E}^2) 
\notag \\ & 
+\frac{1}{P^2+4q_{E}^2} \left( -\frac{14 p_E^4}{q_E^2}+\frac{17 p_E^2 P^2}{2 q_E^2}+96 p_E^2+\frac{7 P^4}{q_E^2}+20 P^2+24 q_E^2 \right) \theta (q_{E}^2-p_{E}^2) ,
\label{BS-kernel-1}
\end{align}
where we have defined
$
C := \frac{Ng^2}{16\pi^2}  .
$
We can rewrite the kernel into the form  which facilitates selecting the dominant terms:
\begin{align}
 & K(p_{E}^2,q_{E}^2;P^2) =  \frac{q_E^2}{P^2+4q_{E}^2}\left(  
18+88\frac{q_E^2}{p_E^2} +10 \frac{P^2}{q_E^2} 
+\frac{37}{2}\frac{P^2  }{p_E^2}
+ 7\frac{P^2}{p_E^2}\frac{P^2}{q_E^2}
\right) \theta (p_{E}^2-q_{E}^2) 
\notag \\ & 
+\frac{q_E^2}{P^2+4q_{E}^2} \left( 
24 +96 \frac{p_E^2}{q_E^2} -14\frac{p_E^4}{q_E^4} 
+20 \frac{P^2}{q_E^2} +7\frac{P^4}{q_E^4}
+\frac{17}{2}\frac{p_E^2 }{ q_E^2}\frac{P^2}{q_E^2}
\right) \theta (q_{E}^2-p_{E}^2) ,
\label{BS-kernel-1b}
\end{align}
\end{subequations}

\subsection{$P^\mu=0$ case}

First, we consider the limit $P^\mu=0$ of (\ref{BS-kernel-1}).
This is a first step to examine the existence of the solution of the BS equation, although this limit is obviously unphysical for the bound state problem. 
This helps us to check whether or not our method works in this problem. 

The BS equation (\ref{BS-kernel-1}) reduces
%
%
\begin{align}
3 \left( {p_{E}^2}+m^2 \right) ^2 f({p_{E}^2})
=& C\Big[ \int _0^{p_{E}^2} dq_{E}^2 \left(  \frac{9q_{E}^2}{2} + \frac{7 q_{E}^4}{4{p_{E}^2}} \right) f(q_{E}^2)
\nonumber\\&
+\int _{{p_{E}^2}}^\infty dq_{E}^2 \left(  6q_{E}^2 +24{p_{E}^2} - \frac{7{p_{E}^4} }{2q_{E}^2} \right) f(q_{E}^2) \Big] .
\label{int_eq_P0}
\end{align}

In order to solve the integral equation, we leave only the term in the integrand which is expected to give the most dominant contribution to the integral, and put the momentum cutoff $\Lambda$ as the upper limit of  integration:
\begin{equation}
  \left( p_{E}^2 + m^2 \right) ^2 f(p_{E}^2)=C \left[ \int _0^{p_{E}^2} dq_{E}^2 \frac{3q_{E}^2}{2} f(q_{E}^2)+\int _{p_{E}^2}^{\Lambda^2} dq_{E}^2 2q_{E}^2 f(q_{E}^2) \right] ,
\label{2.12}
\end{equation}
By defining the function $A({p_{E}^2})$ by
\begin{equation}
A({p_{E}^2}) := ({p_{E}^2}+m^2)^2 f({p_{E}^2}) ,
\end{equation}
we can rewrite (\ref{2.12}) into 
\begin{equation}
 A({p_{E}^2})=C\left[ \int _0^{p_{E}^2} dq_{E}^2 \frac{3q_{E}^2}{2(q_{E}^2+m^2)^2} A(q_{E}^2)+\int _{p_{E}^2}^{\Lambda^2} dq_{E}^2 \frac{2q_{E}^2}{(q_{E}^2+m^2)^2} A(q_{E}^2) \right] .
\label{2.15}
\end{equation}
In (\ref{2.15}), we have IR value at ${p_{E}^2} = 0$:
\begin{equation}
A(0)=2C \int _0^{\Lambda^2} dq_{E}^2 \frac{q_{E}^2}{(q_{E}^2+m^2)^2} A(q_{E}^2) ,
\end{equation}
while we have UV value at ${p_{E}^2} = \Lambda^2$:
\begin{equation}
A(\Lambda^2 )=\frac{3C}{2} \int _0^{\Lambda^2} dq_{E}^2 \frac{q_{E}^2}{(q_{E}^2+m^2)^2} A(q_{E}^2) .
\end{equation}
Therefore, we find that the ratio is given by
\begin{equation}
\frac{A({\Lambda^2} )}{A(0)}=\frac{3}{4} .
\label{ir-bc-r}
\end{equation}

By differentiating both sides of (\ref{2.15}) with respect to ${p_{E}^2}$, we obtain the differential equation which is equivalent to the original integral equation (\ref{2.15}):
\begin{equation}
\frac{d A({p_{E}^2})}{d{p_{E}^2}}=-\frac{C}{2}\frac{{p_{E}^2}}{({p_{E}^2}+m^2)^2}A({p_{E}^2}) .
\label{diff1}
\end{equation}
By  solving this equation, we obtain the general solution and the BS amplitude as  
\begin{align}
A({p_{E}^2})&={\text{Const.}}\times \frac{e^{-\frac{C}{2}\frac{m^2}{{p_{E}^2}+m^2}}}{({p_{E}^2}+m^2)^\frac{C}{2}} ,
\quad
f(p_{E}^2) ={\text{Const.}}\times \frac{e^{-\frac{C}{2}\frac{m^2}{p_{E}^2+m^2}}}{(p_{E}^2+m^2)^{\frac{C}{2}+2}} .
\label{sol-diff1}
\end{align}

For the solution (\ref{sol-diff1}) of the differential equation (\ref{diff1}) to be the solution of the original integral equation, the boundary condition (\ref{ir-bc-r}) must be satisfied:
\begin{align}
  \frac{A({\Lambda^2})}{A(0)}&=  \left(\frac{m^2}{m^2+{\Lambda^2}} \right)^{\frac{C}{2}} e^{ \frac{C}{2}\frac{\Lambda^2}{\Lambda^2+m^2}} 
=\frac{3}{4} ,
\end{align}
which leads to the relation:
\begin{align}
  \ln \frac{m^2}{m^2+{\Lambda^2}} +    \frac{\Lambda^2}{{\Lambda^2} +m^2}  = - \frac{2}{C} \ln \frac{4}{3}  .
\end{align}
When $\Lambda$ is quite large compared with $m$, this relation reduces to 
\begin{equation}
\ln \frac{m^2}{m^2+{\Lambda^2}}= -\gamma, 
\quad 
\gamma := 1+\frac{2}{C}\ln \frac{4}{3}.
\end{equation}
Then the mass $m$ obeys the relation:
\begin{gather}
m^2=\frac{{\Lambda^2}}{e^\gamma-1}   .
\end{gather}
Provided that $C$ becomes $\Lambda$ dependent and that $C\to 0$ as $\Lambda \to \infty$, $\gamma$ is approximated as $\gamma \sim  \frac{2}{C}\ln \frac{4}{3}$ and the gluon mass $m$ obeys the scaling law for large $\Lambda$,
\begin{equation}
 m^2 \sim {\Lambda^2} e^{-\frac{2}{C}\ln \frac{4}{3}} .
\end{equation}
The coupling $C$ must go to zero as $\Lambda \to \infty$ so that $m$ converges to a finite and nonzero value. This is consistent with the (ultraviolet) asymptotic freedom.

\subsection{$P^\mu \not =0$ case}

In what follows, we consider the $P^\mu \not =0$ case.

\subsubsection{$P^\mu \ne 0$ case: analytical study} 

First, we give an analytical treatment. For this purpose, we incorporate the leading terms and neglect the subleading terms among the $P^2$ dependent and $p^2$ dependent terms in the numerator of the kernel $K$ on the right-hand side of (\ref{BS-kernel-1b}) where the denominator of the kernel $K$ is regarded as $P^2+4q_{E}^2$.
Then the BS equation reads 
\begin{align}
3 \left( p_E^2-\frac{P^2}{4}+m^2\right) ^2f(p_E^2;P^2)&=C\Bigg[ \int _0^{p_E^2}q_E^2dq_E^2 \frac{18q_E^2+10P^2}{P^2+4q_E^2} f(q_E^2;P^2) \notag \\
&\hspace{1cm}+\int _{p_E^2}^\infty q_E^2dq_E^2 \frac{24q_E^2+20P^2+\frac{7P^4}{q_E^2}}{P^2+4q_E^2} f(q_E^2;P^2) \Bigg]   .
\label{BSE-kernel2}
\end{align}
The neglected terms are incorporated in the numerical treatment afterwards. 

In the case of $P^\mu \ne 0$, we introduce the reduced variables $x,y,r$ in units of $P^2$ as
\begin{align}
x  := \frac{p_{E}^2}{P^2} \ge 0 , \
y := \frac{q_{E}^2}{P^2}  \ge 0 , \
r := \frac{m^2}{P^2}  \ge 0 .
\end{align}
For the momentum cutoff $\Lambda$ of the Euclidean momentum $p_{E}$, we define the reduced variable $\lambda$ by
\begin{align}
 \lambda := \frac{\Lambda^2}{P^2} > 0 .
\end{align}
In order to consider the two-gluon bound state, we restrict the region of $r$ to
\begin{align}
 r> \frac{1}{4} \Longleftrightarrow P^2 < 4m^2=(2m)^2 .
\end{align}
Then (\ref{BSE-kernel2}) is rewritten as
\begin{equation}
\left( 4x+4r-1\right) ^2 f(x)=\frac{16C}{3}\left[ \int _0^x dy \frac{18y^2+10y}{4y+1} f(y) +\int _x^\lambda dy \frac{24y^2+20y+7}{4y+1} f(y) \right]  .
\label{BSE-bb1}
\end{equation}

Moreover, we introduce the scaled amplitude $A$ by
\begin{equation}
A(x)\equiv \left( x-\frac{1}{4}+r \right) ^2 f(x) .
\label{fA}
\end{equation}
Then the integral equation (\ref{BSE-bb1}) is rewritten in terms of  $A(x)$ as 
\begin{equation}
A(x)=\frac{16C}{3}\left[ \int _0^x dy \frac{18y^2+10y}{(4y+1)(4y+4r-1)^2} A(y) +\int _x^\lambda dy \frac{24y^2+20y+7}{(4y+1)(4y+4r-1)^2}  A(y) \right] .
\label{integ-diff1}
\end{equation}
In order to obtain the differential equation which is equivalent to the integral equation (\ref{integ-diff1}), we differentiate (\ref{integ-diff1}) with respect to $x$. 
Then the equivalent differential equation is given by
\begin{equation}
  \frac{dA(x)}{dx} =-\frac{16C(6x^2+10x+7)}{3(4x+1)(4x+4r-1)^2}A(x) .
\label{de-A-P1}
\end{equation}

The differential equation is solved:
\begin{align}
A(x)&=\frac{{\text{Const.}}}{(4x+1)^{\frac{13C}{8(2r-1)^2}} (4x+4r-1)^{\frac{C(16r^2-16r-9)}{8(2r-1)^2}}} \exp \left[ -\frac{C(48r^2-104r+79)}{12(2r-1)(4x+4r-1)}\right]  .
\end{align}
Thus, we obtain the BS amplitude: 
\begin{align}
f(x)&=\frac{{\text{Const.}}}{(4x+1)^{\frac{13C}{8(2r-1)^2}} (4x+4r-1)^{\frac{C(16r^2-16r-9)}{8(2r-1)^2}+2}} \exp \left[ -\frac{C(48r^2-104r+79)}{12(2r-1)(4x+4r-1)}\right]  .
\label{BS-amplitude-ana}
\end{align}

For a solution of the differential equation to become the solution of the original integral equation, it must satisfy the boundary conditions. 
The boundary conditions are obtained:
IRBC by putting $x \to 0$ in (\ref{integ-diff1}):
\begin{equation}
A(0)=\frac{16C}{3}\int _{0}^{\lambda} dy \frac{24y^2+20y+7}{(4y+1)(4y+4r-1)^2}  A(y) ,
\label{IRbc}
\end{equation}
UVBC by putting $x \to \lambda$ in (\ref{integ-diff1}):
\begin{equation}
A(\lambda )=\frac{16C}{3}\int _{0}^{\lambda} dy \frac{18y^2+10y}{(4y+1)(4y+4r-1)^2} A(y)  .
\label{UVbc}
\end{equation}
By substituting the solution (\ref{BS-amplitude-ana}) into (\ref{IRbc}) or (\ref{UVbc}), we can in principle obtain the relation for $C$, $r$ and $\lambda$, which we call the scaling relation. 
However, the resulting expression in this case will be too complicated to extract the physical results we need. 
Therefore, we move to the numerical investigations. 
\subsubsection{$P^\mu \ne 0$ case: numerical study}

\begin{figure}[t]
\centering
\includegraphics[scale=0.70]{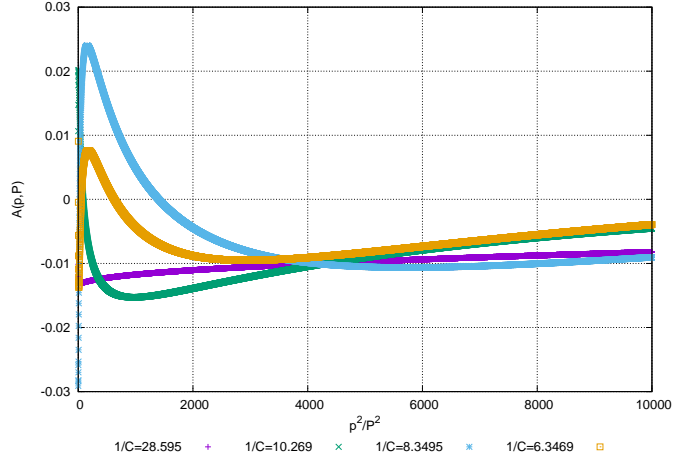}
\caption{
The gluon-gluon BS amplitudes $A$ as a function of $p^2/P^2$ associated to four different eigenvalues  obtained for $\lambda=10000$ and $P^2/m^2 =1$. 
}
\label{fig:BS-amplitude-f1}
\end{figure}


From now on, we proceed to perform the numerical study of the BS equation with the original kernel (\ref{BS-kernel-1}).
We encounter new features once we take into account the $p^2$ dependence of the kernel (\ref{BS-kernel-1}).
In the numerical solution of the BS equation the eigenvalue of $C$ is not uniquely determined, even if the values of $\lambda$ and $r$ are fixed, namely, the ultraviolet cutoff $\Lambda$ and the constituent gluon mass $m$ are given in units of $P^2$. 
In other words, $P^2$ can have a number of possible values for a given coupling constant $C$, even if the ultraviolet cutoff $\Lambda$ and the constituent gluon mass $m$ are fixed. 
These values converge to a set of eigenvalues if the number of partitions for approximating the integral are increased.  

Fig.~\ref{fig:BS-amplitude-f1} shows  the real part of the BS amplitude as a function of $p^2/P^2$ (momentum in units of $P$) for the cutoff $\lambda=10^4$ and a small value $r=1$. 
The plots are given for the gluon-gluon BS amplitudes $A$ associated to four different eigenvalues.   
The resulting BS amplitudes associated to the different eigenvalues have the different number of nodes. 
In general, the eigenvalue and the associated BS amplitude can be complex valued. 
We have confirmed that the numerical solutions for the BS amplitude  are real valued, when the largest eigenvalue is real valued.  In fact, we have checked that the imaginary parts vanish for the numerical solutions. 
The numerical solution of the nodeless BS amplitude with the largest eigenvalue plotted in Fig.~\ref{fig:BS-amplitude-f1} is consistent with the continuum solution (\ref{BS-amplitude-ana}).

\begin{figure}[t]
\centering
\includegraphics[scale=0.6]{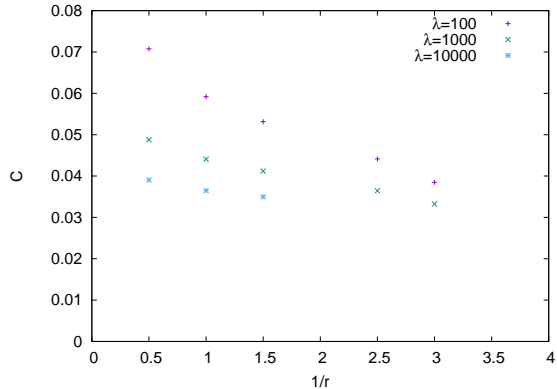}
\vskip -0.3cm
\caption{
The plot of the scaling relation: $C$ vs. $1/r=P^2/m^2$ 
for various values of the cutoff $\lambda$, $\lambda=100,1000,10000$ (from top to bottom). 
This is obtained by solving the BS equation  (\ref{BS-kernel-1}) in a numerical way.   
}
\label{fig:C-1/r-2}
\end{figure}

Fig.~\ref{fig:C-1/r-2} is the plot of the coupling $C$ as a function of the ratio $1/r:=P^2/m^2$ for various but fixed values of the cutoff $\lambda$, which represents the scaling relation. 

\begin{figure}[t]
\centering
\includegraphics[scale=0.6]{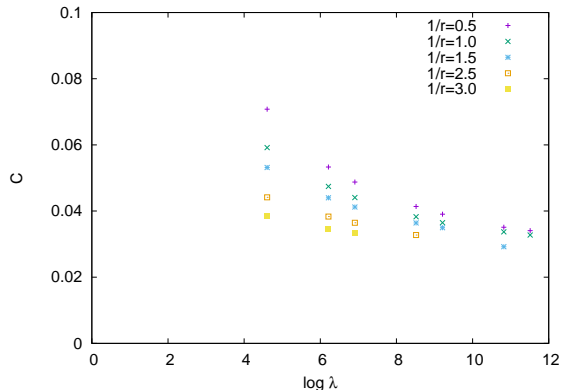}
\vskip -0.5cm
\caption{
The coupling constant $C$ as a function of the momentum scale $\lambda$ for a fixed value of $1/r=P^2/m^2$, which is required for the scaling relation to hold.
The plots from top to bottom correspond to $1/r=P^2/m^2=0.5,1.0,1.5,2.5,3.0$. 
}
\label{fig:running-coupling2}
\end{figure}

Fig.~\ref{fig:running-coupling2} is the plot of  the coupling constant $C$ as a function of the momentum scale $\lambda$ for various but fixed values of $1/r:=P^2/m^2$, which is required for the scaling relation represented by Fig.~\ref{fig:C-1/r-2} to hold.  
This is the renormalization group flow which gives the same physics, $P/m=\text{const.}$, namely, the same bound state mass for a given gluon mass $m$. 
This result will be valid for relatively large $\lambda$ due to the approximations adopted in this paper. 
As the cutoff $\lambda$ increases,  the coupling $C$ decreases like $1/\ln \lambda$ and finally vanishes as $\lambda \to \infty$: 
\begin{align}
 C \cong \frac{\rm{Const.}}{\ln \lambda} .
\end{align}
This behavior is consistent with the (ultraviolet) asymptotic freedom expected to hold in the Yang-Mills theory.

\begin{figure}[h]
\centering
\includegraphics[scale=0.75]{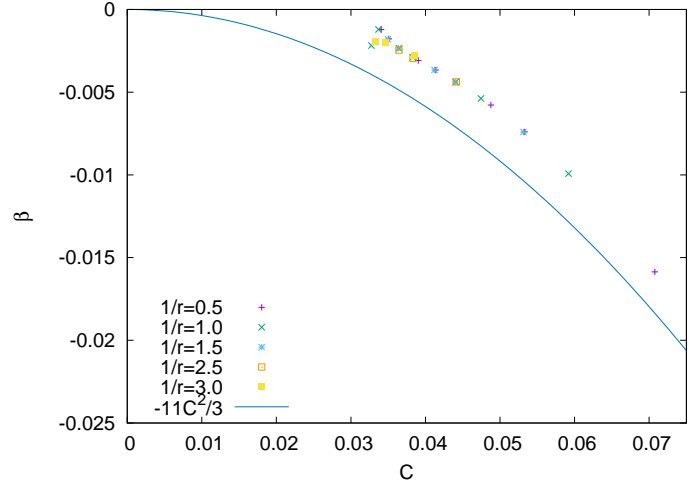}
\vskip -0.3cm
\caption{
The plot of the $\beta$ function $\beta(C):=\lambda \frac{dC}{d\lambda}$  obtained from the numerical solutions of the BS equation (\ref{BS-kernel-1}) for various values of $P/m$: $P^2/m^2=0.5,1.0,1.5,2.5,3.0$. 
The solid line denotes  the $\beta$ function $\beta(C)=-\frac{11}{3}C^2$ obtained  from the perturbation theory to one loop.
}
\label{fig:beta_function3}
\end{figure}

Fig.~\ref{fig:beta_function3} shows the plot of the $\beta$ function $\beta(C):=\lambda \frac{dC}{d\lambda}$ calculated from the running  of the coupling $C$ with respect to $\lambda$  (Fig.~\ref{fig:running-coupling2}) for various values of $P/m$.
We find that all the $\beta$ functions are negative and coincide for small $C$ or large $\lambda$ independently of $P/m$. 
The resulting $\beta$ function can be compared with the beta function $\beta(C)=-\frac{11}{3}C^2$ obtained by one-loop calculation in perturbation theory, which is indicated by the solid curve in Fig.~\ref{fig:beta_function3}.
We find that all the points are well fitted to the single curve, although they are located slightly above the solid curve.  
We observe that the $\beta$ function obtained from the numerical solution of the BS equation has the same sign and  order as the one-loop $\beta$ function.
It should be remarked that the exact agreement of our result with the one loop prediction is not anticipated, since we have neglected some terms already in obtaining the  kernel (\ref{BS-kernel-1}). 
Our results indicate that the bound state solution of the BS equation is consistent  with the asymptotic freedom in the Yang-Mills theory. This is one of the main results of this paper.

Finally, we give a comment how the derivative of $C$ with respect to $\lambda$ is calculated numerically.  The derivative was obtained by taking the finite difference between the smallest eigenvalues of $C$ at $\lambda$ and $1.1\lambda$ for any fixed value of $r^{-1}=P^2/m^2$. 
This approximation for the derivative becomes worse for larger $\lambda$.   
In view of this, we have used the data for the cutoff $\lambda=10^5$ which is the largest possible value we can take due to the limitation of the computer memory available to us. 



\section{Conclusion and discussion}

We summarize the results obtained in this paper. 
First, we have given a systematic derivation of the BS equation for the gluon and ghost BS amplitude by starting from the CJT effective action for the massive Yang-Mills theory.  The solutions of the BS equation represent the simultaneous bound states for gluons and ghosts.  

Second, we have obtained the numerical solutions of the derived BS equation in the ladder approximation with the standing gauge coupling, and the improved ladder approximation with the running gauge coupling.
As a warming up problem, we restricted to the vanishing total momentum  $P^\mu =0$.
The we have found that 
(i) the gluon--gluon contribution is dominant in the solution,  
(ii) the ghost BS amplitude is much smaller  than (at most 10\% of) the gluon BS amplitude, 
(iii) the infrared behavior is influenced by the effect of the running coupling constant.

Third, armed with these preliminary investigations, we proceeded to obtain an approximate solution for the  gluon-gluon BS amplitude alone, which  corresponds to a two-gluon bound state as a candidate for a color singlet scalar glueball. 
We have calculated the beta functions through the running of the coupling constant by using the numerical solutions of the BS equation with an approximated kernel for different choices of the parameters, the ultraviolet cutoff and the constituent gluon mass. 
We have shown that the numerical results converge to a single function which has the same sign and order of magnitude as the one-loop beta function, 
although the exact agreement with the one-loop perturbation theory is not reproduced due to the approximation we adopted in obtaining  the kernel of the BS equation. 
Thus, we have shown that the solution for the gluon bound state obtained in our approximation for the BS equation is consistent with the (ultraviolet) asymptotic freedom of the Yang-Mills theory.  


It is obvious that we have a number of unsolved problems to be tackled:
improvement of the kernel for the BS equation by incorporating more terms, examination of the high-energy behavior for performing the normalization of the BS amplitude, renormalization of the BS equation, improvement of the approximation to keep gauge invariance, e.g., through the Ward-Takahashi relations, confirmation of physical unitarity of the massive Yang-Mills theory, and so on. 
We hope to report some of the results in the near future.



%









\section*{Acknowledgment}

This work is  supported by Grants-in-Aid for Scientific Research (C) No.24540252 and (C) No.15K05042 from the Japan Society for the Promotion of Science (JSPS).


\appendix
\section{BS kernel with a gauge parameter} \label{Appendix_BS_ladder}\label{section:kernel-general}

We decompose the gluon propagator into the transverse part $t_{\mu \nu}$ and the longitudinal part $\ell_{\mu \nu}$:
\begin{align}
D_{\mu \nu}^{(0)ab}(k)&=\frac{-i\delta ^{ab}}{k^2}\left( g_{\mu \nu}-\frac{k_\mu k_\nu}{k^2}\right) +\frac{-i\delta ^{ab}}{k^2} \alpha  \frac{k_\mu k_\nu}{k^2} 
\notag \\
&:= \frac{-i\delta ^{ab}}{k^2} \left[ t_{\mu \nu}(k)+\alpha \ell_{\mu \nu}(k) \right]  . 
\end{align}
Here 
$t_{\mu \nu}(k)$ and $\ell_{\mu \nu}(k)$ satisfy the properties:
\begin{align}
k_\mu t_{\mu \nu}(k) =0 ,
\quad 
k_\mu \ell_{\mu \nu}(k) =  k_\nu .
\end{align}
In order to simplify the expression, we use the identities:
\begin{align}
p^\pm  =k^-+q^\pm =k^+-q^\mp ,
\quad
q^\pm  =-k^-+p^\pm =k^+-p^\mp .
\end{align}
The approximate kernel $\mathcal{K}^{(0)}$ used in this paper is obtained from $\mathcal{K}$ of Fig.~\ref{fig:BS}
by replacing the full propagators by the bare ones and neglecting the loop diagrams in $K$.


The ghost-ghost kernel is given by
\begin{align}
\mathcal{K}^{(0)ab;cd}
&=-\Gamma ^{(0)eca}_\lambda (k^-,q^+,p^+)D^{(0)ee}_{\lambda \beta}(k^-)\Gamma ^{(0)ebd}_\beta (-k^-,p^-,q^-)\Delta ^{(0)cc}(q^+)\Delta ^{(0)dd}(q^-) 
\notag \\
&=-ig^2\frac{f^{eac}f^{ebd}}{( k^-)^2 (q^-)^2 (q^+)^2}\left\{  p_\lambda ^+ t_{\lambda \beta}(k^-)p_\beta ^- +q_\lambda ^+ \alpha \ell_{\lambda \beta}(k^-)p_\beta ^-\right\} . \label{GHGH}
\end{align}



The ghost-gluon kernel is given by
\begin{align}
& \mathcal{K}^{(0)}{}^{ab;cd}_{\ \ ;\mu '\nu '}
\notag \\
=&-\Big( \Gamma ^{(0)cea}_{\lambda}(q^+,k^-,p^+)\Delta ^{(0)ee}(k^-)\Gamma ^{(0)dbe}_{\beta}(-q^-,p^-,k^-)D^{(0)cc}_{\lambda \mu '}(q^+)D^{(0)dd}_{\beta \nu'}(q^-) \notag \\
&\hspace{1cm}+\Gamma ^{(0)dea}_{\lambda}(-q^-,k^+,p^+)\Delta ^{(0)ee}(k^+) \Gamma ^{(0)cbe}_{\beta}(q^+,p^-,k^+)D^{(0)cc}_{\beta \mu '}(q^+)D^{(0)dd}_{\lambda \nu'}(q^-) \Big) \notag \\
%
%
%
%
%
%
=&-ig^2 \frac{f^{eac}f^{ebd}}{(k^-)^2 (q^-)^2 (q^+)^2} \left\{ p_\lambda ^+t_{\lambda \mu '}(q^+)-\alpha q_{\mu '}^+ +p_\lambda ^+ \alpha \ell_{\lambda \mu '}(q^+)\right\} \left\{ p_\beta ^-t_{\beta \nu '}(q^-)+p^-_\beta \alpha \ell_{\beta \nu '}(q^-)\right\} \notag \\
&-ig^2 \frac{f^{ead}f^{ebc}}{(k^+)^2 (q^-)^2 (q^+)^2}\left\{ p_\lambda ^+t_{\lambda \nu '}(q^-)+\alpha q^-_{\nu '}+p^+_\lambda \alpha \ell_{\lambda \nu '}(q^-)\right\} \left\{ p_\beta ^-t_{\beta \mu '}(q^+)+p^-_\beta \alpha \ell_{\beta \mu '}(q^+)\right\} .
\label{GHGL}
\end{align}




The gluon-ghost kernel is given by
\begin{align}
\mathcal{K}^{(0)}{}^{ab;cd}_{\mu \nu ;} =&-\Big[ \Gamma ^{(0)ace}_\mu (-p^+,q^+,-k-)\Delta ^{(0)ee}(-k^-)\Gamma ^{(0)bed}_\nu (p^-,-k^-,q^-)\Delta ^{(0)cc}(q^+)\Delta ^{(0)dd}(q^-) \notag \\
&\hspace{1cm}+\Gamma ^{(0)aed}_\mu (-p^+,k^+,q^-)\Delta ^{(0)ee}(k^+)\Gamma ^{(0)bce}_\nu (p^-,q^+,k^+)\Delta ^{(0)cc}(q^+)\Delta ^{(0)dd}(q^-)\Big] \notag \\
%
%
%
%
=&ig^2\frac{f^{eac}f^{ebd}}{(k^-)^2 (q^-)^2 (q^+)^2}q^+_\mu k^-_\nu -ig^2\frac{f^{ead}f^{ebc}}{(k^+)^2 (q^-)^2 (q^+)^2}k^+_\mu q^+_\nu .
\label{GLGH}
\end{align}
The gluon-ghost kernel $\mathcal{K}^{(0)cd;ab}_{\ \ ;\mu \nu}$ does not have the gauge parameter dependence, since the external gluon propagator does not contribute to the kernel and there does not exist ghosts in the internal lines. 
 

The gluon-gluon kernel $\mathcal{K}^{(0)ab;cd}_{\ \ \, \mu \nu ;\mu '\nu '}$ consists of three parts.
\begin{subequations}
\begin{align}
\mathcal{K}^{(0)}{}^{ab;cd}_{\mu \nu ;\mu '\nu '} = K_{\text{3Gl}}{}^{ab;cd}_{\mu \nu ;\mu '\nu '} + K_{\text{3Gl}}^{\text{cross}}{}^{ab;cd}_{\mu \nu ;\mu '\nu '} + K_{\text{4Gl}}{}^{ab;cd}_{\mu \nu ;\mu '\nu '} ,
\end{align}
\begin{align}
K_{\text{3Gl}}{}^{ab;cd}_{\mu \nu ;\mu '\nu '} &=\Gamma ^{(0)ace}_{\mu \gamma \lambda}(-p^+,q^+,k^-)D^{(0)ee}_{\lambda \lambda '}(k^-) \Gamma ^{(0)bde}_{\nu \gamma ' \lambda '}(p^-,-q^-,-k^-)D^{(0)cc}_{\gamma \mu '}(q^+)D^{(0)dd}_{\gamma '\nu'}(q^-) 
\notag \\
%
%
%
%
&=ig^2\frac{f^{eac}f^{ebd}}{(k^-)^2(q^-)^2(q^+)^2} \left[ t_{\lambda \lambda '}(k^-)+\alpha \ell_{\lambda \lambda '}(k^-) \right] \left[ t_{\gamma '\nu '}(q^-)+\alpha \ell_{\gamma '\nu '}(q^-)\right] 
\notag \\
&\hspace{0cm} \times 
\left[ t_{\gamma \mu '}(q^+)+\alpha \ell_{\gamma \mu '}(q^+)\right] 
\left[ (q^+_\gamma -2p^+_\gamma) g_{\lambda \mu}+(2p^+_\lambda -k^-_\lambda )g_{\gamma \mu}+(p^+_\mu -2q^+_\mu )g_{\lambda \gamma}\right] 
\notag \\
&\hspace{0cm} \times \left[ (2p^-_{\lambda '}-k^-_{\lambda '})g_{\nu \gamma '}+(q^-_{\gamma '}-2p^-_{\gamma '})g_{\nu \lambda '}+(p^-_\nu -2q^-_\nu )g_{\lambda '\gamma '}\right] ,
\end{align}

\begin{align}
K_{\text{3Gl}}^{\text{cross}}{}^{ab;cd}_{\mu \nu ;\mu '\nu '} =&\Gamma ^{(0)ade}_{\mu \gamma \lambda}(-p^+,-q^-,k^+)D^{(0)ee}_{\lambda '\lambda}(k^+)\Gamma ^{(0)bce}_{\nu \gamma '\lambda '}(p^-,q^+,-k^+)D^{(0)cc}_{\gamma '\mu '}(q^+)D^{(0)dd}_{\gamma \nu '}(q^-) 
\notag \\
%
%
%
%
=&ig^2\frac{f^{ead}f^{ebc}}{(k^+)^2(q^-)^2(q^+)^2}\left[ t_{\lambda \lambda '}(k^+)+\alpha \ell_{\lambda \lambda '}(k^+)\right] \left[ t_{\gamma \nu '}(q^-)+\alpha \ell_{\gamma \nu '}(q^-)\right] 
\notag \\
&\hspace{0cm} \times
\left[ t_{\gamma '\mu'}(q^+)+\alpha \ell_{\gamma '\mu'}(q^+) \right] 
  \left[ (2p^+_\lambda -k^+_\lambda )g_{\gamma \mu}-(2p^+_\gamma +q^-_\gamma )g_{\lambda \mu}+(p^+_\mu +2q^-_\mu )g_{\lambda \gamma}\right] 
\notag \\
&\hspace{0cm}\times \left[ (2p^-_{\lambda '}-k^+_{\lambda '})g_{\nu \gamma '}-(2p^-_{\gamma '}+q^+_{\gamma '})g_{\nu \lambda '}+(p^-_\nu +2q^+_\nu )g_{\lambda '\gamma '}\right] ,
\end{align}

\begin{align}
K_{\text{4Gl}}{}^{ab;cd}_{\mu \nu ;\mu '\nu '} &=\Gamma ^{(0)abdc}_{\mu \nu \gamma \delta}(-p^+,p^-,-q^-,q^+)D^{(0)cc}_{\delta \mu '}(q^+)D^{(0)dd}_{\gamma \nu '}(q^-) \notag \\
&=\frac{g^2}{(q^-)^2(a^+)^2} \left[ t_{\gamma \nu '}(q^-)+\alpha \ell_{\gamma \nu '}(q^-)\right] \left[ t_{\delta \mu}(q^+)+\alpha \ell_{\delta \mu}(q^+)\right] 
\notag \\
&\hspace{0cm}\times  [ f^{ead}f^{ebc}\left( g_{\gamma  \delta}g_{\mu  \nu}-g_{\gamma  \nu}g_{\delta  \mu} \right) +f^{eac}f^{ebd}\left( g_{\gamma  \delta}g_{\mu  \nu}-g_{\gamma \mu}g_{\delta  \nu}\right) 
\notag \\
&-f^{eab}f^{ecd}\left( g_{\gamma  \mu}g_{\delta \nu}-g_{\gamma  \nu}g_{\delta  \mu}\right) ] .
\end{align}
\end{subequations}




\section{BS Kernel in the Landau gauge}\label{section:kernel-Landau}

\subsection{$P^\mu \not = 0$ }\label{Landau_BSkernel}

In the Landau gauge ($\alpha =0$), the kernels given in the above take the simpler forms:
\begin{align}
\mathcal{K}^{(0)ab;cd}=&-ig^2\frac{f^{eac}f^{ebd}}{(k^-)^2(q^-)^2(q^+)^2}\left( p^+\cdot p^- +\frac{(p^+\cdot k^-)(k^-\cdot p^-)}{(k^-)^2}\right) ,
\end{align}
\begin{align}
\mathcal{K}^{(0)ab;cd}_{\ \ \ \ \ ;\mu '\nu '}=&-ig^2\frac{f^{eac}f^{ebd}}{(k^-)^2(q^-)^2(q^+)^2}\left( p^+_{\mu '}-q^+_{\mu '}\frac{p^+ \cdot q^+}{(q^+)^2}\right) \left( p^-_{\nu '}-q^-_{\nu '}\frac{p^-\cdot q^-}{(q^-)^2}\right) \notag \\
&-ig^2\frac{f^{ead}f^{ebc}}{(k^+)^2(q^-)^2(q^+)^2}\left( p^-_{\mu '}-q^+_{\mu '}\frac{p^-\cdot q^+}{(q^+)^2}\right) \left( p^+_{\nu '}-q^-_{\nu '}\frac{p^+\cdot q^-}{(q^-)^2}\right) ,
\end{align}
\begin{align}
\mathcal{K}^{(0)ab;cd}_{\ \ \, \mu \nu ;}=&ig^2\frac{f^{eac}f^{ebd}}{(k^-)^2 (q^-)^2 (q^+)^2}q^+_\mu k^-_\nu -ig^2\frac{f^{ead}f^{ebc}}{(k^+)^2 (q^-)^2 (q^+)^2}k^+_\mu q^+_\nu ,
\end{align}
\begin{align}
\mathcal{K}^{(0)ab;cd}_{\ \ \, \mu \nu ;\mu '\nu '}=&ig^2 \frac{f^{eac}f^{ebd}}{(k^-)^2(q^-)^2(q^+)^2} \notag \\
&\hspace{0cm} \times \Big[ -4p^-_{\lambda '}t_{\mu \lambda '}(k^-)t_{\nu \nu '}(q^-)t_{\gamma \mu '}(q^+)p^+_\gamma +4p^+_\lambda t_{\lambda \lambda '}(k^-)p^-_{\lambda '}t_{\nu \nu '}(q^-)t_{\mu \mu '}(q^+) 
\notag \\
&\hspace{0.5cm} +2p^+_\mu t_{\gamma \lambda '}(k^-)p^-_{\lambda '}t_{\nu \nu '}(q^-)t_{\gamma \mu '}(q^+) +4t_{\mu \nu}(k^-)p^-_{\gamma '}t_{\gamma '\nu '}(q^-)t_{\gamma \mu '}(q^+)p^+_\gamma \notag \\
&\hspace{0.5cm} -4p^+_\lambda t_{\lambda \nu}(k^-)p^-_{\gamma '}t_{\gamma '\nu '}(q^-)t_{\mu \mu '}(q^+)-2p^+_\mu t_{\gamma \nu}(k^-) p^-_{\gamma '}t_{\gamma '\nu '}(q^-)t_{\gamma \mu '}(q^+) 
\notag \\
&\hspace{0.5cm} +(p^-_\nu -2q^-_\nu )\big\{ -2t_{\mu \lambda '}(k^-)t_{\lambda '\nu '}(q^-)t_{\gamma \mu '}(q^+)p^+_\gamma  
\notag \\
&\hspace{0.5cm}
+2p^+_\lambda t_{\lambda \lambda '}(k^-)t_{\lambda '\nu '}(q^-)t_{\mu \mu '}(q^+)
+p^+_\mu t_{\gamma \lambda '}(k^-)t_{\lambda '\nu '}(q^-)t_{\gamma \mu '}(q^+)\big\} \Big]
\notag \\
&+ig^2 \frac{f^{eac}f^{ebd}}{(k^+)^2(q^-)^2(q^+)^2} \notag \\
&\hspace{0cm} \times \Big[ 4p^+_\lambda t_{\lambda \lambda '}(k^+)p^-_{\lambda '}t_{\mu \nu '}(q^-)t_{\nu \mu '}(q^+)-4t_{\mu \lambda '}(k^+)p^-_{\lambda '}p^+_\gamma t_{\gamma \nu '}(q^-)t_{\nu \mu '}(q^+) \notag \\
&\hspace{0.5cm}  +2(p^+_\mu +2q^-_\mu )p^-_{\lambda '}t_{\lambda \lambda '}(k^+)t_{\lambda \nu '}(q^-)t_{\nu \mu '}(q^+) -4p^+_\lambda t_{\lambda \nu}(k^+)t_{\mu \nu '}(q^-)t_{\gamma '\mu '}(q^+)p^-_{\gamma '} \notag \\
&\hspace{0.5cm}  +4t_{\mu \nu}(k^+)p^+_\gamma t_{\gamma \nu '}(q^-)p^-_{\gamma '}t_{\gamma '\mu '}(q^+)-2(p^+_\mu +2q^-_\mu )t_{\lambda \nu}(k^+)t_{\lambda \nu '}(q^-)p^-_{\gamma '}t_{\gamma '\mu '}(q^+) \notag \\
&\hspace{0.5cm}  +(p^-_\nu +2q^+_\nu )\big\{ 2p^+_\lambda t_{\lambda \lambda '}(k^+)t_{\mu \nu '}(q^-)t_{\lambda '\mu '}(q^+)-2t_{\mu \lambda '}(k^+)p^+_\gamma t_{\gamma \nu '}(q^-)t_{\lambda '\mu '}(q^+) \notag \\
&\hspace{0.5cm}+(p^+_\mu +2q^-_\mu )t_{\lambda \lambda '}(k^+)t_{\lambda \nu '}(q^-)t_{\lambda '\mu '}(q^+) \big\} \Big] 
\nonumber
\end{align}
\begin{align}
&-ig^2 \frac{1}{(q^-)^2(q^+)^2}\Big[ f^{ead}f^{ebc}\left( t_{\gamma \nu '}(q^-)t_{\gamma \mu '}(q^+)g_{\mu  \nu}-t_{\nu \nu '}(q^-)t_{\mu \mu '}(q^+) \right) \notag \\
&\hspace{3cm}+f^{eac}f^{ebd}\left( t_{\gamma \nu '}(q^-)t_{\gamma \mu '}(q^+)g_{\mu  \nu}-t_{\mu \nu '}(q^-)t_{\nu \mu '}(q^+)\right) \notag \\
&\hspace{3cm}-f^{eab}f^{ecd}\left( t_{\mu \nu '}(q^-)t_{\nu \mu '}(q^+)-t_{\nu \nu '}(q^-)t_{\mu \mu '}(q^+)\right) \Big] .
\end{align}

\subsection{$P^\mu =0$ }\label{section:kernel-Landau_P=0}

We give the kernel in the simplest case $P_\mu =0$.
The BS amplitudes obtained by solving the BS equation at  $P_\mu =0$ from the beginning may agree with those obtained in the limit of $P_\mu =0$ of the solution of the BS equation with $P_\mu \not =0$, if the vector model has the same property as the Wick--Cutkosky model which is the exactly solved model for the scalar particles. 
Notice that transversality changes at $P_\mu =0$.  In the limit  $P_\mu \to 0$,  $A_2$ is ill-defined.
The ordinary transverse tensor is reproduced at $P_\mu =0$, since there is no difference between $k^+$ and $k^-$ at $P_\mu =0$
The BS equation cannot contain $A_2$ part once $P_\mu =0$ is adopted: 
\begin{align}
\begin{pmatrix}
A_1(p) \\ B(p)
\end{pmatrix}
=&-ig^2\int \frac{d^4q}{(2\pi )^4}\frac{f^{eac}f^{ead}}{(p-q)^2 q^4}
\begin{pmatrix}
a_{11} & a_{12} \\
a_{21} & a_{22} \\
\end{pmatrix}
\begin{pmatrix}
A_1(q) \\ B(q)
\end{pmatrix} \notag \\
&-ig^2\int \frac{d^4q}{(2\pi )^4}\frac{f^{eac}f^{ead}}{(p+q)^2 q^4}
\begin{pmatrix}
b_{11} & b_{12} \\
b_{21} & 0 \\
\end{pmatrix}
\begin{pmatrix}
A_1(q) \\ B(q)
\end{pmatrix} .
\end{align}
Here $a_{ij}$ denotes the contribution from the non-crossing diagrams, while $b_{ij}$ from the crossing diagrams.
Therefore, $b_{22}=0$ is due to the fact that there do not exist the ghost crossing diagrams in the ghost-ghost kernel.  
The explicit form of the matrix elements are as follows.
\begin{align}
a_{11}^{\text{3Gl}}&=\frac{14 p^4 q^2+p^2 \left( -38 q^2 p\cdot q-11 (p\cdot q)^2+35 q^4\right) +4\left( -9 q^4 p\cdot q+q^2 (p\cdot q)^2+5 (p\cdot q)^3+3 q^6\right)}{3 q^2\left( p-q\right) ^2} ,
\nonumber\\
a_{11}^{\text{4Gl}}&=-6(p-q)^2 ,
\nonumber\\
a_{12}&=\frac{1}{3}\left( p\cdot q-q^2\right) ,
\quad
a_{21} =-\frac{p^2 q^2-(p\cdot q)^2}{2q^2} ,
\quad
a_{22} =-\frac{p^2 q^2-(p\cdot q)^2}{(p-q)^2} ,
\nonumber\\
b_{11}&=\frac{14 p^4 q^2+p^2 \left( 38 q^2 p\cdot q-11 (p\cdot q)^2+35 q^4\right) +4\left( 9 q^4 p\cdot q+q^2 (p\cdot q)^2-5 (p\cdot q)^3+3 q^6\right)}{3 q^2 \left( p+q\right) ^2} ,
\nonumber\\
b_{12}&=-\frac{1}{3}\left( p\cdot q+q^2\right) ,
\quad
b_{21} =-\frac{p^2 q^2-(p\cdot q)^2}{2q^2} .
\end{align}
We perform the Wick rotation to $p,q$.  
\begin{align}
p_0&=ip_4 ,
\quad
p^2 =p_0^2-p_1^2-p_2^2-p_3^2=-(p_1^2+p_2^2+p_3^2+p_4^2):= -p_E^2 .
\end{align}
This is applied also to $q$.
In what follows, we omit the index $E$. 
Consequently, we can perform the integration over the four-dimensional coordinates by converting it to the integration in the four-dimensional Euclidean space. 
\begin{align}
\begin{pmatrix}
A_1(p) \\ B(p)
\end{pmatrix}
=&g^2\int \frac{q^2\sin ^2\theta dq^2d\theta}{(2\pi )^3}\frac{f^{eac}f^{ead}}{(p-q)^2 q^4}
\begin{pmatrix}
a_{11} & a_{12} \\
a_{21} & a_{22} \\
\end{pmatrix}
\begin{pmatrix}
A_1(q) \\ B(q)
\end{pmatrix} \notag \\
&+g^2\int \frac{q^2\sin ^2\theta dq^2d\theta}{(2\pi )^3}\frac{f^{eac}f^{ead}}{(p+q)^2 q^4}
\begin{pmatrix}
b_{11} & b_{12} \\
b_{21} & 0 \\
\end{pmatrix}
\begin{pmatrix}
A_1(q) \\ B(q)
\end{pmatrix} \notag \\
=&g^2f^{eac}f^{ead}\int \frac{dq^2}{16\pi ^2}\frac{2}{\pi}\int \frac{q^2\sin ^2\theta d\theta}{(p-q)^2q^4}
\begin{pmatrix}
a_{11} & a_{12} \\
a_{21} & a_{22} \\
\end{pmatrix}
\begin{pmatrix}
A_1(q) \\ B(q)
\end{pmatrix} \notag \\
&+g^2f^{eac}f^{ead}\int \frac{dq^2}{16\pi ^2}\frac{2}{\pi}\frac{q^2\sin ^2\theta d\theta}{(p+q)^2 q^4}
\begin{pmatrix}
b_{11} & b_{12} \\
b_{21} & 0 \\
\end{pmatrix}
\begin{pmatrix}
A_1(q) \\ B(q)
\end{pmatrix} .
\end{align}
Assuming that $A_1(q)$ and $B(q)$ do not depend on the angles, we can perform the angular integration to obtain
\begin{align}
\begin{pmatrix}
A_1(p^2) \\ B(p^2)
\end{pmatrix}
=&g^2f^{eac}f^{ead}\int \frac{dq^2}{16\pi ^2}
\begin{pmatrix}
a_{11} & a_{12} \\
a_{21} & a_{22} \\
\end{pmatrix}
\begin{pmatrix}
A_1(q^2) \\ B(q^2)
\end{pmatrix} \notag \\
&+g^2f^{eac}f^{ead}\int \frac{dq^2}{16\pi ^2}
\begin{pmatrix}
b_{11} & b_{12} \\
b_{21} & 0 \\
\end{pmatrix}
\begin{pmatrix}
A_1(q^2) \\ B(q^2)
\end{pmatrix} ,
\end{align}
where the matrix elements are given as 
\begin{align}
a_{11}&=a_{11}^{3Gl}+a_{11}^{4Gl} \notag \\
&=\left\{ \left( \frac{11}{3 p^2}+\frac{15}{4 q^2}\right) +\left( -\frac{6}{q^2}\right) \right\} \theta (p^2-q^2)+\left\{ \left( -\frac{7 p^4}{12 q^6}+\frac{4 p^2}{q^4}+\frac{4}{q^2}\right) +\left( -\frac{6}{q^2}\right) \right\} \theta (q^2-p^2) ,
\nonumber\\
a_{12}&=-\frac{1}{6 p^2}\theta (p^2-q^2)+\frac{1}{6} \frac{p^2-2 q^2}{q^4} \theta (q^2-p^2) ,
\nonumber\\
a_{21}&=\frac{1}{8} \left(\frac{1}{p^2}-\frac{3}{q^2}\right) \theta (p^2-q^2)+\frac{p^4-3 p^2 q^2}{8 q^6} \theta (q^2-p^2) ,
\nonumber\\
a_{22}&=\left( -\frac{3}{4 p^2}\right) \theta (p^2-q^2)+\left( -\frac{3 p^2}{4q^4}\right) \theta (q^2-p^2) ,
\nonumber\\
b_{11}&=\left( \frac{11}{3 p^2}+\frac{15}{4q^2}\right) \theta (p^2-q^2)+\left( -\frac{7 p^4}{12 q^6}+\frac{4 p^2}{q^4}+\frac{4}{q^2}\right) \theta (q^2-p^2) ,
\nonumber\\
b_{12}&=\left( -\frac{1}{6 p^2}\right) \theta (p^2-q^2)+\left( \frac{p^2-2 q^2}{6 q^4} \right) \theta (q^2-p^2) ,
\nonumber\\
b_{21}&=\frac{1}{8} \left(\frac{1}{p^2}-\frac{3}{q^2}\right) \theta (p^2-q^2) +\frac{p^4-3 p^2 q^2}{8 q^6} \theta (q^2-p^2) .
\end{align}
Here we have used the formulae for the angular integrations which are given e.g., in \cite{Fukamachi15}. 
We find that the non-crossing and crossing diagrams give the same contributions to the kernel. 
This is consistent with the assumption that $A_1$ and $B$ depend only on the magnitude of the Euclidean momentum. 
\begin{equation}
\begin{pmatrix}
A_1(p^2) \\ B(p^2)
\end{pmatrix}
=C\int dq^2
\begin{pmatrix}
\tilde{a}_{11} & \tilde{a}_{12} \\
\tilde{a}_{21} & \tilde{a}_{22} \\
\end{pmatrix}
\begin{pmatrix}
A_1(q^2) \\ B(q^2)
\end{pmatrix} ,
\label{5.40}
\end{equation}
with
\begin{align}
\tilde{a}_{11}&=\left\{ \frac{22}{3 p^2}+\frac{3}{2 q^2}\right\} \theta (p^2-q^2)+\left\{ -\frac{7 p^4}{6 q^6}+\frac{8 p^2}{q^4}+\frac{2}{q^2}\right\} \theta (q^2-p^2) ,
\nonumber\\
\tilde{a}_{12}&=\left\{ -\frac{1}{3 p^2}\right\} \theta (p^2-q^2)+\left\{ \frac{p^2-2 q^2}{3 q^4}\right\} \theta (q^2-p^2) ,
\nonumber\\
\tilde{a}_{21}&=\left\{ \frac{1}{4} \left(\frac{1}{p^2}-\frac{3}{q^2}\right) \right\} \theta (p^2-q^2)+\left\{ \frac{p^4-3 p^2 q^2}{4 q^6}\right\} \theta (q^2-p^2) ,
\nonumber\\
\tilde{a}_{22}&=\left\{ -\frac{3}{4 p^2}\right\} \theta (p^2-q^2)+\left\{ -\frac{3 p^2}{4 q^4}\right\} \theta (q^2-p^2) .
\label{5.44}
\end{align}
Here we have defined 
$C:= \frac{g^2N}{16\pi ^2}$ using $\frac{g^2}{16\pi ^2}f^{eac}f^{ead}=\frac{g^2N}{16\pi ^2}\delta ^{cd}$ for $SU(N)$.




\end{document}